\documentclass[12pt]{article}
\usepackage{epsfig}
\usepackage{psfrag}
\usepackage{enumitem}
\usepackage{latexsym}
\usepackage{indentfirst}
\usepackage{fancyhdr}
\usepackage{amssymb}
\usepackage{amsmath}
\usepackage{amsfonts}
\usepackage{colordvi}
\usepackage{pifont}
\usepackage{cite}
\usepackage{color}
\usepackage[footnotesize]{caption2}
\usepackage{graphicx}
\usepackage[center,footnotesize,hang]{subfigure}
\usepackage{url}
\usepackage{array}

\def\beq{\begin{equation}}
\def\eeq{\end{equation}}
\def\be{\begin{equation}}
\def\ee{\end{equation}}
\def\bea{\begin{eqnarray}}
\def\eea{\end{eqnarray}}
\def\wt{\widetilde}
\def\ol{\overline}

\def\vev#1{\langle #1 \rangle}

\begin{document}

\title{\hfill ~\\[-30mm]
       \hfill\mbox{\small IPPP-12-56}\\[-3mm]
       \hfill\mbox{\small DCPT-12-112}\\[14mm]
       \textbf{A Grand $\boldsymbol{\Delta(96)\times SU(5)}$ Flavour Model}} \date{}
\author{\\[2mm]
Stephen F. King$^{1\,}$\footnote{E-mail: king@soton.ac.uk}~,~~ 
Christoph Luhn$^{2\,}$\footnote{E-mail: christoph.luhn@durham.ac.uk}~,~~
Alexander J. Stuart$^{1\,}$\footnote{E-mail: a.stuart@soton.ac.uk}
\\[8mm]
  \emph{$^1$\small{}School of Physics and Astronomy, University of Southampton,}\\ \emph{\small Southampton, SO17 1BJ, U.K.}\\[4mm]
 \emph{$^2$\small{}Institute for Particle Physics Phenomenology, University of Durham,}\\  \emph{\small Durham, DH1 3LE, U.K.}}
 
\maketitle

\vskip 0.5cm

\begin{abstract}
\noindent
Recent results from the Daya Bay and RENO reactor experiments have measured the smallest 
lepton mixing angle and found it to have a value of
$\theta_{13}\approx9^{\circ}$.  This result presents a new challenge for the
existing paradigms of discrete flavour symmetries which attempt to describe
all quark and lepton masses and mixing angles.
Here we propose a Supersymmetric Grand Unified Theory of Flavour
based on $\Delta(96)\times SU(5)$, together with a $U(1)\times Z_3$ symmetry, including 
a full discussion of $\Delta(96)$ in a convenient basis. The Grand $\Delta(96)\times SU(5)$ Flavour Model
relates the quark mixing angles and masses in the form of the
Gatto-Sartori-Tonin relation and realises the Georgi-Jarlskog mass relations
between the charged leptons and down-type quarks. We predict a
Bi-trimaximal (not Tri-bimaximal) form of neutrino mixing matrix,
which, after including charged lepton corrections with zero phase,
leads to the following GUT scale predictions for the 
atmospheric, solar, and reactor mixing angles:
$\theta_{23}\approx 36.9^{\circ}$, 
$\theta_{12}\approx 32.7^{\circ}$ and 
$\theta_{13}\approx 9.6^{\circ}$, in good agreement with recent
global fits, and a zero Dirac CP phase $\delta \approx 0$. 
\end{abstract}
\thispagestyle{empty}
\vfill
\newpage
\setcounter{page}{1}

%%%%%%%%%%%%%%%%
\newpage
%%%%%%%%%%%%%%%%

\section{Introduction}
\label{Introduction}

It is one of the goals of theories of particle physics beyond the
Standard Model to predict quark and lepton masses and mixings,
or at least to relate them. While the quark mixing angles are known to all be
rather small, by contrast two of the lepton mixing angles, the atmospheric
angle $\theta_{23}$ and the solar angle $\theta_{12}$, are identified as being
rather large~\cite{Nakamura:2010zzi}.  Until recently the remaining reactor
angle $\theta_{13}$ was unmeasured. Direct evidence for $\theta_{13}$ was
first provided by T2K, MINOS and Double
Chooz~\cite{Abe:2011sj,Adamson:2011qu,Abe:2011fz}. 
Recently Daya Bay~\cite{DayaBay}, RENO~\cite{RENO}, and Double
Chooz~\cite{DCt13} Collaborations have measured $\sin^2(2\theta_{13})$:
\begin{eqnarray}
\label{t13}
\begin{array}{cc}
\text{Daya Bay: } & \sin^2(2\theta_{13})=.089\pm.011 \text{(stat.)}\pm.005
\text{(syst.)}\ ,\\
\text{RENO: }    & \sin^2(2\theta_{13})= .113\pm.013\text{(stat.)}\pm.019
\text{(syst.)\ ,}\\
\text{Double Chooz: }& \sin^2(2\theta_{13})=.109\pm.030\text{(stat.)}\pm.025
\text{(syst.)}\ .\\
\end{array}
\end{eqnarray}

From a theoretical or model building point of view, one significance of this
measurement is that it excludes the tri-bimaximal (TB) lepton mixing
pattern~\cite{TBM} in which the atmospheric angle is maximal, the reactor angle
vanishes, and the solar mixing angle is approximately $35.3^{\circ}$. 
When comparing global fits to TB mixing it is convenient to express the solar,
atmospheric and reactor angles in terms of deviation parameters ($s$, $a$ and
$r$) from TB mixing~\cite{rsa}: 
\be
\label{rsadef}
\sin \theta_{12}=\frac{1}{\sqrt{3}}(1+s),\ \ \ \ 
\sin\theta_{23}=\frac{1}{\sqrt{2}}(1+a), \ \ \ \ 
\sin \theta_{13}=\frac{r}{\sqrt{2}}.
\ee
The global fit in \cite{global} 
(which has been updated to include the data released at the Neutrino 2012
Conference; see also \cite{global2})
yields the $1\sigma$ ranges for the TB deviation parameters:
\begin{eqnarray}
\label{rsafit}
\begin{array}{ccc}
-0.066\leq s\leq -0.013,&-0.146\leq a\leq -0.094,& 0.208\leq r\leq 0.231,
\end{array}
\end{eqnarray}
assuming a normal neutrino mass ordering.
As well as showing that TB is excluded by the reactor angle being non-zero, 
Eq.~(\ref{rsafit}) shows a preference for the atmospheric angle to
be below its maximal value and also a slight preference for the solar angle to
be below its tri-maximal value. 

As a result of the rapidly changing landscape of neutrino mixing parameters,
many models based on discrete family symmetry (see~\cite{Reviews} and
references therein) which were proposed initially to account for TB mixing are
now either excluded, or have been subjected to
modification~\cite{theta13before,theta13after}.   
A promising new approach has emerged based on a new discrete family symmetry,
namely $\Delta(96)$, which is capable of predicting the value of the 
reactor mixing angle \cite{Delta(96)orig1,DingDelta(96)}. It was found that
$\Delta(96)$ can lead to two alternative mixing patterns, related by exchange
of the lower two rows of the PMNS mixing matrix, depending  on the particular
choice of Klein symmetry $Z^S_2\times Z^U_2$ that is respected by the
neutrino mass matrix
without re-ordering the three generations of lepton doublets.
In this paper we shall select the following Klein group generators 
for the triplet representation ${\bf 3}$ of $\Delta(96)$ (see
Appendix~\ref{Appendix}):  
\begin{eqnarray}
S= \frac{1}{3} \left(
\begin{array}{ccc}
 -1 & 2 & 2 \\
 2 & -1 & 2 \\
 2 & 2 & -1
\end{array}\right), \;
U=\frac{1}{3} \left(
\begin{array}{ccc}
 -1+\sqrt{3} & -1-\sqrt{3} & -1 \\
 -1-\sqrt{3} & -1 & -1+\sqrt{3} \\
 -1 & -1+\sqrt{3} & -1-\sqrt{3}
\end{array}
\right) , \label{SandU}
\end{eqnarray}
where $SU=US$ and $S^2=U^2=I$.  
The breaking of the $\Delta(96)$ family
symmetry to this particular Klein symmetry in the neutrino sector will be
achieved through a set of scalar fields (flavons) coupling to
neutrino mass terms.  These flavons will obtain vacuum expectation values
(vevs) that will only leave invariant the $Z^S_2\times Z^U_2$ subgroup
contained in $\Delta(96)$. 

Assuming the above choice of Klein symmetry in the neutrino sector, 
together with a diagonal charged lepton mass matrix,
$\Delta(96)$ predicts the following PMNS mixing matrix:
\begin{eqnarray}
\label{Unu0}
U_{\rm{BT}}=\left(
\begin{array}{ccc}
 a_+ & \frac{1}{\sqrt{3}} & a_- \\
 -\frac{1}{\sqrt{3}} & \frac{1}{\sqrt{3}} & \frac{1}{\sqrt{3}} \\
a_- & -\frac{1}{\sqrt{3}} & a_+
\end{array}
\right)P,
\end{eqnarray}
where $a_{\pm}=(1\pm \frac{1}{\sqrt{3}})/2$ and
$P$ is the usual diagonal
matrix containing the Majorana phases. We shall refer to this as
``Bi-trimaximal'' (BT) mixing due to the distinctive St George's cross 
feature of the middle row and column being of the tri-maximal form.\footnote{An alternative choice of Klein symmetry leads to the
second and third rows of Eq.~(\ref{Unu0})
being interchanged \cite{Delta(96)orig1}, and hence an atmospheric angle
in the second octant, somewhat disfavoured by the recent global fits.}
This leads to the following predictions: 
\begin{eqnarray}
\label{delta96pred2}
\begin{array}{l}
\sin \theta_{12} =\sin \theta_{23}= \sqrt{\frac{8-2\sqrt{3}}{13}}\approx 0.591 
\ \  \ \ (\theta_{12}=\theta_{23}\approx 36.2^{\circ}), \\
\sin \theta_{13}= a_-\approx 0.211 
\ \  \ \ (\theta_{13}\approx 12.2^{\circ}).
\end{array}
\end{eqnarray}
In terms of the TB deviations, the BT mixing pattern predicts:
 \begin{eqnarray}
 \label{delta96pred3}
\begin{array}{ccc}
s\approx 0.023,& a\approx -0.165,& r\approx 0.299, 
\end{array}
\end{eqnarray}
which all fall outside the $1\sigma$ ranges for these parameters in Eq.~\eqref{rsafit}.
This motivates going beyond the simple models of leptons proposed so far
\cite{DingDelta(96)}, and in particular to Grand Unified models where the charged lepton
mass matrix is only approximately diagonal and the resulting charged lepton mixing 
corrections will slightly modify the above predictions, bringing them into agreement with
the global fits. 

The purpose of the present paper is to construct the first $\Delta(96)\times
SU(5)$ Grand Unified Theory (GUT) of Flavour in a Supersymmetric (SUSY)
context, where the model leads to BT neutrino mixing, modified by charged
lepton corrections. The model relates the quark mixing angles and masses in
the form of the Gatto-Sartori-Tonin (GST)~\cite{GST} relation and realises the
Georgi-Jarlskog (GJ)\cite{GJ} mass relations between the charged leptons and
down-type quarks. 
In order to do this we have to develop the group theory of $\Delta(96)$ beyond that
which appears in \cite{DingDelta(96)} where two models of leptons were
proposed based on $\Delta(96)$. The reason is that in~\cite{DingDelta(96)} the
charged lepton mass eigenvalues required re-ordering before a physical
interpretation could be achieved.  However such a re-ordering is not
convenient from the point of view of GUT theories, since the swapping of rows
and columns can cloud the hierarchical structures that one wishes to achieve. 
Here we calculate the Clebsch-Gordan (CG) coefficients in a suitable basis
right from the start. We also emphasise the use of the $S$, $T$, and $U$
generators to draw analogy to previous models based on $S_4$ ($A_4$). The
lengthy but necessary group theoretical aspect of this work is relegated to
the Appendix. 

The remainder of this work is as follows. In Section \ref{Model}, the
$\Delta(96)\times SU(5)$ SUSY GUT is introduced by defining its fields and
their transformation properties under $\Delta(96)$, $SU(5)$, as well as an
additional $U(1)$ and $Z_3$ symmetry.  The resulting mass matrices are
calculated after flavour and electroweak symmetry breaking to reveal the quark
and lepton mass matrices from which the predictions for the quark and lepton  
mixing parameters are obtained. 
In Section~\ref{Vacuum Alignment} the vacuum alignments
assumed in Section~\ref{Model} are justified by constructing a flavon
potential and deriving the flavon alignments from minimisation conditions.  In Section \ref{NLO}, 
possible subleading corrections to the leading order (LO) predictions are discussed.
Section~\ref{Conclusion} concludes the paper. The group theory of $\Delta(96)$ is elicited in
Appendix~\ref{Appendix} by constructing its generators, character table, and
corresponding CG coefficients.

%%%%%%%%%%%%%%%%%%%%%%%%%%%%%%%%%%%%%%

%%%%%%%%%%%%%%%%%%%%%%%%%%%%%%%%%%%%%%

%%%%%%%%%%%%%%%%%%%%%%%%%%%%%%%%%%%%%%

\section{The $\boldsymbol{\Delta(96)\times SU(5)}$ Model }
\label{Model}

\subsection{Fields, Symmetries and Yukawa Operators}
In this section, we present an $SU(5)$ GUT, endowed with a $\Delta(96)$
flavour symmetry (the relevant group theory of $\Delta(96)$ can be found in
Appendix A).  We follow closely the logic of the $S_4\times SU(5)$ GUT
presented in Ref.~\cite{S4GUT}.  Therefore, we assign the $\bf{\overline{5}}$
matter fields of $SU(5)$ to the triplet representation ${\bf 3}$ of
$\Delta(96)$ and call them $F$.  The first two particle generations' ${\bf
  10}$ dimensional matter fields are assigned to the doublet ${\bf 2}$ of $\Delta(96)$,
and the third generation's ${\bf 10}$ dimensional matter field is assigned to
the singlet ${\bf 1}$ (call these $T$ and $T_3$, respectively).  Also,
right-handed neutrino fields, $N$, are added that transform as a singlet under
$SU(5)$ and a ${\bf \overline{3}}$ under $\Delta(96)$.  

In addition to the above 
matter fields, the GUT Higgs fields, $H_5$, $H_{\overline{5}}$, and
$H_{\overline{45}}$ are added and are all singlets under the $\Delta(96)$
flavour symmetry.  Notice that the MSSM Higgs doublets $H_u$ and $H_d$ (whose
vevs are related by $v_u/v_d=\tan \beta$)  originate from $H_5$ and a linear
combination of $H_{\overline{5}}$ and $H_{\overline{45}}$,
respectively.\footnote{The orthogonal linear combination of $H_{\overline{5}}$
and $H_{\overline{45}}$ is assumed to decouple from the low-energy theory by
obtaining a GUT scale mass. Note that the $SU(5)$ Higgs potential will not be
studied because the aim of 
this work is to construct a theory of flavour.  Therefore, minimising the full GUT
Higgs potential is beyond the scope of this work.}  We have added the
$H_{\overline{45}}$ to obtain the GJ relations between the charged lepton and
down quark masses.    
With this minimal field content transforming under $SU(5)$ and $\Delta(96)$,
the only Yukawa terms in the superpotential which are allowed are $FNH_5$,
$T_3T_3H_5$, and $TTH_5$.  These renormalisable terms correspond to a Dirac
neutrino mass, a top quark mass and masses for up and charm quarks,
respectively.  
The former term is problematic because it produces a degenerate $\sim 100$ GeV
neutrino mass spectrum (if Yukawa couplings are not tuned accordingly) with no
mixing.  The latter term is problematic 
because it yields degenerate masses for the up and the charm quark,
and provides little explanation for the observed patterns of quark mixing.

In view of the above phenomenological requirements, 
it is necessary to introduce new symmetries which forbid any
un-phenomenological operator, as well as to add more fields
(\textit{i.e.} flavon fields) which couple to the existing matter fields to
fix the problematic predictions mentioned above.
The additional flavon fields, denoted by $\Phi^f_\rho$, are all $SU(5)$
  gauge singlets and charged under the family symmetry.
Here $f$ represents the fermion sector that the flavon field couples to and can
be either $u$, $d$, or $\nu$, and $\rho$ labels the representation of
$\Delta(96)$ that the flavon field transforms under.  Thus, $\rho$ can
take any of the ten values of irreducible representations of $\Delta(96)$
(\textit{i.e.} ${\bf{1}}$, ${\bf{1^{\prime}}}$, ${\bf{2}}$, ${\bf{3}}$,
${\bf{\widetilde{3}}}$, ${\bf{\overline{3}}}$, ${\bf{3^{\prime}}}$,
${\bf{\widetilde{3}^{\prime}}}$, ${\bf{\overline{3}^{\prime}}}$, and
${\bf{6}}$). 
In addition to these flavon fields, a global $U(1)$ symmetry must be
introduced to prevent a flavon field associated with one fermionic sector
contaminating another sector, as well as preventing the leading order $TTH_5$ 
term.  This $U(1)$ symmetry will not be gauged in order to avoid the
constraints of anomaly cancellation.  However, it will be spontaneously broken
by the flavon fields acquiring vevs in our model.  This will lead to a massless
Goldstone boson, unless the $U(1)$ symmetry is also explicitly broken.  For
this reason, the $U(1)$ symmetry is assumed to be explicitly broken in the
hidden sector of the theory so that the Goldstone boson becomes a
Pseudo-Goldstone boson with mass around 1 TeV.  Also, an additional 
$Z_3$ symmetry will be added to prevent the proliferation of problematic terms
in the flavon superpotential (see Section \ref{Vacuum Alignment} for further
explanation).  A summary of the model's field content and its transformation
properties under $\Delta(96)\times SU(5)\times U(1)\times Z_3$ can be found
in Table~\ref{trans}.

\begin{table}[t]
\centering
{\small{
\begin{tabular}{|c||c|c|c|c|c|c||c|c|c|c|c|c|c|c|}\hline
$\!$Field$\phantom{\Big|}\!\!\!$& $T_3$& $T$ & $F$ & $N$ & $H_{5,\overline{5}}$& $H_{\overline{45}}$& $\Phi^u_2$& $\bar{\Phi}^u_2$ & $\Phi_{\overline{3}}^d$ & $\bar{\Phi}_{\overline{3}}^d$ & $\Phi_2^d$ & $\Phi^{\nu}_{\overline{3}^{\prime}}$ & $\Phi^{\nu}_{\widetilde{3}^{\prime}}$&$\Phi^{\nu}_{\widetilde{3}}$\\\hline
$\!\!\!SU(5)$$\phantom{\Big|}\!\!\!\!\!$ & $\bf{10}$ & $\bf{10}$ & $\bf{\overline{5}}$ & $\bf{1}$& $\bf{5}$,$\bf{\overline{5}}$&  $\bf{\overline{45}}$ & $\bf{1}$ &$\bf{1}$&$\bf{1}$&$\bf{1}$&$\bf{1}$&$\bf{1}$&$\bf{1}$&$\bf{1}$\\\hline 
$\!\Delta(96)$$\phantom{\Big|}\!\!\!$ & $\bf{1}$  &$\bf{2}$ & $\bf{3}$ & $\bf{\overline{3}}$ & $\bf{1}$& $\bf{1}$  & $\bf{2}$ & $\bf{2}$&  $\bf{\overline{3}}$ & $\bf{\overline{3}}$& $\bf{2}$ & $\bf{\overline{3}^{\prime}} $ & $\bf{\widetilde{3}^{\prime}}$&$\bf{\widetilde{3}}$\\\hline
$U(1)$$\phantom{\Big|}\!\!$& 0 & $x$ & $y$ & $-y$& 0 & $z$ &$-2x$ & 0 &$-y$&$\!-x-y-2z\!$&$z$&$2y$&$2y$&$w$\\\hline
$Z_3$$\phantom{\Big|}\!\!$  & 1 & 1   &$\omega^2$ & $\omega$ & 1,$\omega$ & $\omega$&$1$&$1$&$1$&$1$&$1$&$\omega$&$\omega$&$\omega$\\\hline
\end{tabular}}}
\caption{\label{trans}The field content of the $\Delta(96)\times SU(5)\times
  U(1)\times  Z_3$ model.  $x$, $y$, $z$, and $w$ are integers, while~$\omega=e^{\frac{2\pi i}{3}}$.}
\end{table}

Using the transformation properties given in Table~\ref{trans}, the lowest
order Yukawa operators invariant under all symmetries can be constructed.  The
resulting terms contributing to the up quark mass matrix read 
\begin{eqnarray}
\label{wu}
y_u^{} T_3T_3H_5+y_u'\frac{1}{M}TT\Phi^u_2H_5+y_u''\frac{1}{M^2}TT\Phi_2^u\bar{\Phi}_2^uH_5,
\end{eqnarray}
where $M$ is a generic messenger scale (around the GUT scale) common to all
higher-dimensional operators and (presumably) order one coupling constants are
denoted by $y_u^{}$, $y_u'$, $y_u''$.  If not explicitly stated, all contractions
of flavon fields are taken into account.  With the up sector written, the
charges in Table~\ref{trans} can be used to write down the Yukawa operators
for the down quarks and charged leptons:
\begin{eqnarray}
\label{wd}
y_d^{}\frac{1}{M}FT_3\Phi^d_{\overline{3}}H_{\overline{5}}+y_d'\frac{1}{M^2}(F\bar{\Phi}^d_{\overline{3}})_1(T\Phi^d_2)_1H_{\overline{45}}+y_d''\frac{1}{M^3}(F\Phi_2^d\Phi_2^d)_3 (T\bar{\Phi}_{\overline{3}}^d)_{\overline{3}} H_{\overline{5}},
\end{eqnarray}
where the subscripts on the parentheses denote the specific contraction being
taken from the tensor product contained inside of the parentheses, and we have
again denoted order one constants by $y_d^{}$, $y_d'$, $y_d''$. Obviously,
there are other contractions involving the fields in parentheses, which are
not written down.  Thus, a specific set of mediator fields existing above the
messenger scale, $M$, which will select out the desired contractions in
Eq.~(\ref{wd}) is assumed. 
For a similar example of how this can be realised in an ultraviolet-completed
  model, see e.g. \cite{S4GUT,Hagedorn:2012ut}.
Notice that the second term in Eq.~(\ref{wd}) gives rise to the GJ relation
(when suitable vevs are applied) which provides phenomenologically acceptable
charged lepton and down quark masses, when extrapolated to low-energy scales.
When the flavon fields in the third term of Eq.~(\ref{wd}) assume suitable
vevs, the GST relation is fulfilled because of equal (12) and (21) entries and
a (11) entry equal to zero, at this order.  This relation gives rise to the
successful prediction of the ratio of the down and strange quark masses to the
Cabibbo angle (\textit{i.e.}
$\theta^q_{12}\approx\theta^d_{12}\approx\sqrt{m_d/m_s}$). Now that the lowest
order quark and charged lepton Yukawa operators are written, attention is
turned to the neutrino sector which has the following leading terms,
constructed from the transformation properties in Table \ref{trans}: 
\begin{eqnarray}
\label{wnu}
y_DFNH_5+\ol y_M NN\Phi^{\nu}_{\overline{3}^{\prime}}+\wt y_M NN\Phi^{\nu}_{\widetilde{3}^{\prime}},
\end{eqnarray}
where we have included the coupling constants $y_D$ and $\ol y_M$, $\wt y_M$
of the Dirac and the Majorana terms, respectively.
We remark that the (auxiliary) flavon field $\Phi^\nu_{\wt {\bf 3}}$ does not couple
to any matter field due to its $U(1)$ charge of $w$. It has been 
introduced to the model for the sole purpose of aligning the other neutrino
flavons, as will be discussed in detail in Section~\ref{Vacuum Alignment}.

%%%%%%%%%%%%%%%%%%%%%%%%%%%%%%%%%%%%%%%%%%%%%%%%%%%%%%%%%%

%%%%%%%%%%%%%%%%%%%%%%%%%%%%%

%%%%%%%%%%%%%%%%%%%%%%%%%%%%%

\subsection{The Quark and Charged Lepton Mass Matrices}
Just as we began the discussion of the invariant Yukawa operators under the
$\Delta(96)\times U(1)\times Z_3$ flavour symmetry by writing the terms contributing to the up-type quark masses first, we begin the discussion of fermion mass matrices by considering the up sector first, as well.  The flavons that must be considered to do this are $\Phi_2^u$ and $\bar{\Phi}_2^u$.  They assume the following vevs (a detailed discussion of the origin of the alignment of these vevs appears in Section 3):
\begin{eqnarray}
\label{uflavvevs} 
\langle\Phi_2^u\rangle=\varphi_2^u
\left(
\begin{array}{c}
0\\
1
\end{array}\right)~\text{ and  }~\langle\bar{\Phi}_2^u\rangle=\bar{\varphi}_2^u
\left(
\begin{array}{c}
0\\
1
\end{array}\right).
\end{eqnarray}
This alignment gives rise to 
\begin{eqnarray}
\label{upmass}
M_u\approx v_u\left(
\begin{array}{ccc}
y_u''\bar{\varphi}_2^u\varphi_2^u/M^2&0&0\\
0&y_u'\varphi_2^u/M&0\\
0&0&y_u^{}
\end{array}\right),
\end{eqnarray}
where $v_u$ denotes the vev of the electroweak Higgs field $H_u$.  Assuming
$\varphi_2^u/M\approx\lambda^4$ and $\bar{\varphi}_2^u/M\approx \lambda^4$,
where $\lambda\approx 0.225$ is the Wolfenstein parameter~\cite{Nakamura:2010zzi} 
associated with the sine of the Cabibbo angle, yields the well-known mass
hierarchy among the up quarks: 
\begin{equation}
m_u:m_c:m_t\approx\lambda^8:\lambda^4:1.
\end{equation}

Moving to the down-type quarks, the flavons $\Phi_2^d$, $\Phi^d_{\overline{3}}$ and $\bar{\Phi}^d_{\overline{3}}$ need to be considered. Assume they acquire the following vacuum alignments:
\begin{eqnarray}
\label{vevphi2d}
\langle\Phi_2^d\rangle=\varphi_2^d
\left(
\begin{array}{c}
1\\
0
\end{array}\right),~~\langle\Phi_{\overline{3}}^d\rangle=\varphi_{\overline{3}}^d
\left(
\begin{array}{c}
0\\
0\\
1
\end{array}\right)\text{,~ and ~}\langle\bar{\Phi}_{\overline{3}}^d\rangle=\bar{\varphi}_{\overline{3}}^d
\left(
\begin{array}{c}
0\\
1\\
-1
\end{array}\right).
\end{eqnarray}
Adopting the left-right convention, these vevs give rise to
the following down quark and charged lepton mass matrices: 
\begin{eqnarray}
\label{dmass}
M_d\approx v_d\left(
\begin{array}{ccc}
0&y_d''(\varphi_2^d)^2\bar{\varphi}^d_{\overline{3}}/M^3   & - y_d''(\varphi_2^d)^2\bar{\varphi}^d_{\overline{3}}/M^3\\
y_d''(\varphi_2^d)^2\bar{\varphi}^d_{\overline{3}}/M^3    &  y_d'\varphi_2^d\bar{\varphi}^d_{\overline{3}}/M^2 -y_d''(\varphi_2^d)^2\bar{\varphi}^d_{\overline{3}}/M^3    & -y_d'\varphi_2^d\bar{\varphi}^d_{\overline{3}}/M^2\\
0  & 0 & y_d^{}\varphi^d_{\overline{3}}/M
\end{array}\right),
\end{eqnarray}
and 
\begin{eqnarray}
\label{emass}
M_e\approx v_d\left(
\begin{array}{ccc}
0&y_d''(\varphi_2^d)^2\bar{\varphi}^d_{\overline{3}}/M^3   &  0\\
y_d''(\varphi_2^d)^2\bar{\varphi}^d_{\overline{3}}/M^3    &  -3y_d'\varphi_2^d\bar{\varphi}^d_{\overline{3}}/M^2 -y_d''(\varphi_2^d)^2\bar{\varphi}^d_{\overline{3}}/M^3    & 0\\
-y_d''(\varphi_2^d)^2\bar{\varphi}^d_{\overline{3}}/M^3  & 3y_d'\varphi_2^d \bar{\varphi}^d_{\overline{3}}/M^2 & y_d^{}\varphi^d_{\overline{3}}/M
\end{array}\right),
\end{eqnarray}
where $v_d$ denotes the vev of the electroweak Higgs field, $H_d$.  
We remark that $M_e$ is obtained from $M_d$ by transposition and
inclusion of the GJ factor of $-3$~\cite{GJ}. The equality\footnote{It is
interesting to note that the (12) and (21) entries of $M_d$ have the same
sign, whereas the signs are opposite in the $S_4\times SU(5)$ model in
Ref.~\cite{S4GUT}. This is a direct consequence of $\Delta(96)$'s CG
coefficients (see Appendix~\ref{Appendix}).}  of the (12) and (21) entries of
$M_d$, together with the vanishing (11) entry at leading order, generates the
desired GST relation, \textit{i.e.}
$\theta^q_{12}\approx\theta^d_{12}\approx\sqrt{m_d/m_s}$~\cite{GST}.

Assuming the following magnitudes for the flavons associated with the down
quark/charged lepton sectors: 
\begin{eqnarray} 
\label{tanbeta}
\begin{array}{ccc}
\varphi^d_{2}/M\approx\lambda\ ,~~&\varphi^d_{\overline{3}}/M\approx\lambda^{1+k}\ ,~~&\bar{\varphi}^d_{\overline{3}}/M\approx\lambda^{2+k}\ ,
\end{array}
\end{eqnarray}
where $k=0$ or $k=1$, the ``GUT scale'' down quark and charged lepton mass
hierarchies may be expressed as: 
\begin{eqnarray}
&& m_e/m_d = 1/3, \ \    m_{\mu}/m_s=3, \ \  m_{\tau}/m_b= 1\ ,\\
&& m_d:m_s:m_b \approx   \lambda^4: \lambda^2:1,
\end{eqnarray}
and the mixing angles $\theta_{ij}^e$ and $\theta_{ij}^d$ take the
semi-familiar forms 
\begin{eqnarray}
\label{downleppred}
\begin{array}{ccc}
\theta_{12}^d\approx \lambda\ ,~&~\theta_{23}^d\approx
\lambda^2\ ,~&~\theta_{13}^d\approx \lambda^3\ ,\\
\theta_{12}^e\approx \lambda/3\ ,~&~\theta_{23}^e\approx
0\ ,~&~\theta_{13}^e\approx 0\ .
\end{array}
\end{eqnarray}

The GJ predictions have been scrutinised in \cite{Ross:2007az}
where it was shown that they are acceptable assuming particular SUSY threshold corrections,
and for particular values of $\tan \beta$. We note that, for the purposes of predicting PMNS mixing angles,
the only aspect of the GJ relations that will be relevant is $\theta_{12}^e\approx \lambda/3$,
where such a charged lepton mixing angle can also be achieved consistently with some of the alternatives 
to the GJ relations proposed by Antusch and Spinrath in \cite{Ross:2007az}, for example
those that predict $m_{\mu}/m_s=9/2$.

Notice that the main difference between the cases $k=0$ and $k=1$ is the mass of the bottom quark and $\tau$ lepton:
\begin{equation}
m_b\approx m_{\tau}\approx\lambda^{1+k}v_d.
\end{equation}
The two different choices for $k$ represent two different predicted ranges for $\tan{\beta}$.  For $k=0$,  $m_b\approx m_{\tau}\approx\lambda v_d$.  While for $k=1$, $m_b\approx m_{\tau}\approx\lambda^2v_d$.  Thus, $k=0$ prefers a larger value of $\tan\beta$ because $\tan\beta$ must make up for the rest of the suppression to the bottom/tau mass.  Therefore, $\tan\beta\approx \lambda^{-2} \approx 25$.  Yet, for $k=1$ the flavons contribute more to the suppression of the mass and $\tan\beta\approx\lambda^{-1}\approx 5$.  For the remainder of this work, we take $k=1$.  This will simplify the number of terms needed to be considered when minimising the flavon potential.  It also allows for a complete listing of the schematic structures of the mass matrices in Eqs.~(\ref{upmass}), (\ref{dmass}), and (\ref{emass}):
\begin{eqnarray}
\label{masssup}
\begin{array}{ccc}
M_u\sim v_u\left(\begin{array}{ccc}
\lambda^8&0&0\\
0&\lambda^4&0\\
0&0&1
\end{array}\right),&
M_d\sim v_d\left(\begin{array}{ccc}
0&\lambda^5&\lambda^5\\
\lambda^5&\lambda^4&\lambda^4\\
0&0&\lambda^2
\end{array}\right),
&M_e\sim v_d\left(\begin{array}{ccc}
0&\lambda^5&0\\
\lambda^5&3\lambda^4&0\\
\lambda^5&3\lambda^4&\lambda^2
\end{array}\right).
\end{array}
\end{eqnarray}
With the reporting of these structures, the discussion of the quark and charged lepton masses and mixings is complete.  The next step is to perform similar arguments for the neutrinos, thereby calculating their masses and mixings at the GUT scale to LO.

%%%%%%%%%%%%%%%%%%%%%%%%%%%%%%%

%%%%%%%%%%%%%%%%%%%%%%%%%%%%%%%

%%%%%%%%%%%%%%%%%%%%%%%%%%%%%%%

\subsection{The Neutrino Mass Matrices }

The aim of this section is the calculation of the neutrino
masses and mixings, associated with a $\Delta(96)$ flavour symmetry.  To this
end, we begin by writing the neutrino Dirac mass matrix given by the Dirac
mass term in Eq.~(\ref{wnu}): 
\begin{eqnarray}
\label{neutdirac}
M_D=y_Dv_u\left(
\begin{array}{ccc}
1&0&0\\
0&1&0\\
0&0&1
\end{array}\right).
\end{eqnarray}
This simple result is due to the fact that no flavons couple to the $FNH_5$
term (see Table \ref{trans}).  However, $\Phi^{\nu}_{\overline{3}^{\prime}}$
and $\Phi^{\nu}_{\widetilde{3}^{\prime}}$ couple to the right-handed Majorana
mass terms; in fact, without these flavons no Majorana masses would exist.
The vacuum alignments of these flavon fields are determined by
demanding that $\Delta(96)$ is broken to the low-energy $Z_2^S\times Z_2^U$
subgroup in the neutrino sector.  Thus, their vevs must be invariant when
acted upon by the corresponding Klein group generators. It is important
to note that these generators take different forms for different $\Delta(96)$
representations. Their relation to the generators given in Eq.~\eqref{SandU}
can be found in Eq.~\eqref{irreps}. With this in mind, we can determine the
vevs of the neutrino flavons in the $\bf{\bar{3}^{\prime}}$  and
$\bf{\widetilde{3}^{\prime}}$ representations which respect the desired Klein
symmetry. We find
\begin{eqnarray}
\label{neutvevs}
\langle\Phi_{\overline{3}^{\prime}}^{\nu}\rangle=\varphi_{\overline{3}^{\prime}}^{\nu}
\left(
\begin{array}{c}
1\\
1\\
1
\end{array}\right),\text{~ and ~}
\langle\Phi_{\widetilde{3}^{\prime}}^{\nu}\rangle=\varphi_{\widetilde{3}^{\prime}}^{\nu}
\left(
\begin{array}{c}
v_1\\
\frac{1}{2}(v_1+v_3)\\
v_3
\end{array}\right),
\end{eqnarray}
where the neutrino flavon vevs
$\varphi_{\overline{3}^{\prime}}^{\nu}$ and $\varphi_{\widetilde{3}^{\prime}}^{\nu}$
are both assumed to be of order $\lambda^4 M$,
in order to bring $M$ (around the GUT scale) down to the seesaw scale of $\sim 10^{13}$ GeV.  Then, using these invariant vevs in the contractions of the relevant irreducible
representations associated with the Majorana mass terms in Eq.~(\ref{wnu})
yields the following leading order contributions to the neutrino Majorana mass
matrix: 
\begin{eqnarray}
\label{neutmajorana}
M_{Maj}=\ol y_M \varphi_{\overline{3}^{\prime}}^{\nu}\left(
\begin{array}{ccc}
 -2  & 1& 1 \\
 1 & -2  & 1\\
 1 & 1 & -2 
\end{array}
\right)+\wt y_M
\varphi_{\widetilde{3}^{\prime}}^{\nu}\left(
\begin{array}{ccc}
 v_3 & v_1 & \frac{1}{2} (v_1+v_3) \\
 v_1 & \frac{1}{2} (v_1+v_3) & v_3 \\
 \frac{1}{2} (v_1+v_3) & v_3 & v_1
\end{array}
\right).
\end{eqnarray}
Using the mass matrices associated with the Dirac [Eq.~(\ref{neutdirac})] and
Majorana [Eq.~(\ref{neutmajorana})] mass terms, the heavy degrees of freedom
associated with the right-handed neutrinos can be integrated out to generate
the seesaw formula~\cite{seesaw}: 
\begin{equation}
m_{\nu}=M_DM_{Maj}^{-1}M_D^T=y_D^2v_u^2M_{Maj}^{-1}.
\end{equation}
Thus, it suffices to diagonalise $m_{\nu}$ to reveal the light neutrino masses.  This can be done with a unitary transformation such that:
\begin{equation}
\label{mnudef}
V_{\nu}m_{\nu} V_{\nu}^T=m_{\nu}^{diag}.
\end{equation}
This straightforward diagonalisation takes the following form:
\begin{eqnarray}
\label{diagmajinv}
m_{\nu}=V_{\nu}^{\dagger}\mathrm{\,diag}(m_1,m_2,m_3)V_{\nu}^{*},
\end{eqnarray}
where the complex masses are given by
\begin{eqnarray}
\begin{array}{ccc}
m_1=\frac{2  y_D^2v_u^2}{\sqrt{3}\wt y_M \varphi_{\widetilde{3}^{\prime}}^{\nu}(v_3-v_1)-6\ol y_M\varphi_{\overline{3}^{\prime}}^{\nu}}, ~  & 
m_2=\frac{2
  y_D^2v_u^2}{3\wt y_M\varphi_{\widetilde{3}^{\prime}}^{\nu}(v_1+v_3)}, ~ &
m_3=\frac{2 y_D^2v_u^2}{\sqrt{3}\wt y_M\varphi_{\widetilde{3}^{\prime}}^{\nu}(v_1-v_3)-6\ol y_M\varphi_{\overline{3}^{\prime}}^{\nu}}, 
\end{array}\label{complmass}
\end{eqnarray}
and
\begin{eqnarray}
\label{Unu}
V_{\nu}^{\dagger}=
\begin{pmatrix}1&0&0\\0&1&0\\0&0&-1
\end{pmatrix}
\left(
\begin{array}{ccc}
 a_+ & \frac{1}{\sqrt{3}} & a_- \\
 -\frac{1}{\sqrt{3}} & \frac{1}{\sqrt{3}} & \frac{1}{\sqrt{3}} \\
a_- & -\frac{1}{\sqrt{3}} & a_+
\end{array}
\right) ,
\end{eqnarray}
with $a_{\pm}=(1\pm \frac{1}{\sqrt{3}})/2$. Up to Majorana phases and the
phase matrix on the left (which will turn out to be unphysical), this is just the BT
mixing form discussed in the Introduction, here applied to the neutrino
sector only. It is easy to see that $V_{\nu}^{\dagger}$ translates to the following
values of the neutrino mixing angle: 
\beq
\label{neutangles}
\theta^{\nu}_{23}=\theta^{\nu}_{12}=\tan ^{-1}\left(\sqrt{3}-1\right)\approx 36.2^{\circ}, 
\ \ \ \  \theta^{\nu}_{13}=\sin ^{-1}(a_-) \approx 12.2^{\circ}.
\eeq

Before calculating the mixing matrix including the charged lepton corrections,
it is worth mentioning that the complex masses in Eq.~\eqref{complmass} do not
satisfy a sum rule in general. This is due to the fact that one of the two
neutrino type flavons entering in Eq.~\eqref{wnu} depends on two free
parameters, $v_1$ and $v_3$, cf. Eq.~\eqref{neutvevs}. However, we will see in
Section~\ref{Vacuum Alignment} that these parameters will be related via the
minimisation conditions of the flavon potential. Anticipating the result of
Eq.~\eqref{unique4} we obtain $v_1=1$ and $v_3=\omega^{2p}$, with
$p=1,2$. It is easy to see that this special relation gives rise to the
following new sum rule of complex neutrino masses:
\be
\frac{1}{m_3} +  \frac{2 i (-1)^p}{m_2} - \frac{1}{m_1} ~=~ 0 \ ,
\ee
where both choices of $p$ allow for either normal or inverted mass ordering.

\subsection{Predictions for PMNS Mixing}
Since the mixing matrices associated with the neutrinos and charged leptons
are calculated, an approximate value for $U_{PMNS}$ elements can be obtained
by first observing that the only leading order nonzero mixing angle in the
charged lepton sector is  $\theta^e_{12}\approx \lambda/3$,
cf. Eq.~\eqref{downleppred}.  In this subsection we consider the effect of such charged
lepton corrections to the lepton mixing angles. 

We begin with a reminder of the conventions used throughout this work.  As
defined in Eq.~(\ref{mnudef}), $V_{\nu}m_{\nu} V_{\nu}^T=m_{\nu}^{diag}$,
where the masses are still complex. The Majorana phase matrix~$P$ which
renders the neutrino masses real and positive is therefore added separately in
the expression for $U_{PMNS}$ in Eq.~\eqref{PMNSconv}. 
Similarly, $V_em_em_e^{\dagger}V_e^{\dagger}=(m_e^{diag})^2$.  Thus, the
left-right convention is adopted for defining the mass matrices.  As a result
of these conventions, 
\begin{equation}
\label{PMNSconv}
U_{PMNS}=V_eV_{\nu}^{\dagger} P .
\end{equation}
As was seen in Eq.~(\ref{downleppred}), $\theta_{23}^e\approx 0$,
$\theta_{13}^e\approx 0$, and $\theta_{12}^e\approx \lambda/3$.  This non-zero
prediction for $\theta_{12}^e$ implies that $V_e$ takes the form, 
\begin{eqnarray}\label{Ve}
V_{e}\approx 
P'\left(\begin{array}{ccc}
\!c^e_{12}& -s^e_{12}e^{-i\delta^e_{12}}&0\!\\
\!s^e_{12}e^{i\delta^e_{12}}&c^e_{12} &0\!\\
\!0&0&1\!
\end{array}
\right),
\end{eqnarray}
where $c_{12}^e =\cos\theta_{12}^e$ and $s_{12}^e =\sin\theta_{12}^e$.
$\delta^e_{12}$ is an undetermined phase and $P'$ is a diagonal matrix
consisting of three unphysical phases which may be absorbed into the charge
lepton mass eigenstates.

Inserting Eqs.~\eqref{Unu} and \eqref{Ve} into Eq.~\eqref{PMNSconv}, we find,
\begin{eqnarray}\label{Eq:PMNS}
U_{\mathrm{PMNS}} \approx 
P''\left(\begin{array}{ccc}
a_+c^e_{12} + \frac{1}{\sqrt{3}}  s^e_{12}e^{-i\delta^e_{12}}
&
 \frac{1}{\sqrt{3}}c^e_{12} - \frac{1}{\sqrt{3}}  s^e_{12}e^{-i\delta^e_{12}}
&
a_-c^e_{12} - \frac{1}{\sqrt{3}}  s^e_{12}e^{-i\delta^e_{12}}
\!\\
a_+s^e_{12}e^{i\delta^e_{12}} - \frac{1}{\sqrt{3}} c^e_{12}
&
\frac{1}{\sqrt{3}}s^e_{12}e^{i\delta^e_{12}} + \frac{1}{\sqrt{3}} c^e_{12} 
&
a_-s^e_{12}e^{i\delta^e_{12}} + \frac{1}{\sqrt{3}} c^e_{12}
\!\\
a_-
&
-  \frac{1}{\sqrt{3}} 
&
a_+
\end{array}
\right)P ,
\end{eqnarray}
where we have permuted the phase matrix $\mathrm{diag}(1,1,-1)$ of
$V^\dagger_\nu$ with the 1-2 rotation of $V_e$ and combined both unphysical
phase matrices into $P''$.

Given $U_{PMNS}$, we identify the reactor angle from
\beq
\sin\theta_{13}=\big|(U_{\mathrm{PMNS}})_{13}\big|\approx \big|a_-c^e_{12} - \mbox{$\frac{1}{\sqrt{3}}$}  s^e_{12}e^{-i\delta^e_{12}}\big|
\approx a_- - \mbox{$\frac{1}{\sqrt{3}}$}  \theta^e_{12}\cos \delta^e_{12} . 
\eeq
In order to reduce the reactor angle from its BT value of $12.2^{\circ}$ down
to the values observed by Daya Bay and RENO we need to assume $\cos
\delta^e_{12}\approx 1$ and hence $\delta^e_{12}\approx 0$, leading to, 
\beq
\sin\theta_{13}
\approx 0.167,
\eeq
corresponding to $\theta_{13}\approx 9.6^{\circ}$.

With the phase $\delta^e_{12}\approx 0$ 
fixed by the requirement of lowering the reactor angle to an acceptable value,
the rest of the angles are easy to read off the matrix, since the PMNS matrix 
in Eq.~\eqref{Eq:PMNS} is then real (up to Majorana phases in $P$) and automatically
in the PDG form with the unphysical phase matrix $P''$ set equal to the unit matrix.
The Dirac CP phase $\delta$ is given by,
\beq
\delta = 
-\arg \big[ (U_{PMNS})_{13}\big]
\approx - \arg \big[ a_- c^e_{12} 
-\mbox{$ \frac{1}{\sqrt{3}}$}  s^e_{12}
\big]
\approx 0 .
\eeq
The atmospheric angle is given by,
\beq
\tan \theta_{23}\approx\frac{\frac{1}{\sqrt{3}}c^e_{12} +a_- s^e_{12}}{a_+}\approx
0.750 ,
\eeq
leading to $\theta_{23}\approx 36.9^{\circ}$, close to the uncorrected value
of $36.2^{\circ}$. 
The solar angle is given by,
\beq
\tan \theta_{12}\approx\frac{ \frac{1}{\sqrt{3}}c^e_{12} - \frac{1}{\sqrt{3}}  s^e_{12}}
{a_+c^e_{12} + \frac{1}{\sqrt{3}}  s^e_{12}}
\approx  0.642 ,
\eeq
leading to $\theta_{12}\approx 32.7^{\circ}$.
It is worth noting that the phase of the solar angle correction is the same as the phase
of the reactor angle correction, so both angles are nicely lowered together
into the desired range.

It is convenient to express the above predictions for the solar, atmospheric
and reactor angles in terms of 
deviation parameters ($s$, $a$ and $r$) from TB mixing
defined in Eq.~\eqref{rsadef} \cite{rsa}:
 \begin{eqnarray}
 \label{delta96pred4}
\begin{array}{ccc}
s\approx -0.065,& a\approx -0.151,& r\approx 0.237. 
\end{array}
\end{eqnarray}
These predictions for $s$, $a$ and $r$ are now in much better agreement
with the $1\sigma$ ranges given in Eq.~\eqref{rsafit}.
This shows that, including charged lepton corrections arising from a GUT model
involving the GJ and GST relations, corrects the BT predictions almost
perfectly, providing that the charged lepton corrections  carry a zero phase.
In particular, the solar angle lies within the $1\sigma$ range, 
while the atmospheric angle almost falls inside the $1\sigma$ allowed interval. 
The predicted reactor angle of $\theta_{13}\approx 9.6 ^{\circ}$ is now
well within the $2\sigma$ range.
However it should also be noted that the above predictions are valid at the GUT scale
and are subject to renormalisation group (RG)~\cite{Antusch:2007ib} and canonical
normalisation (CN)~\cite{CN} corrections. These
effects were neglected since they are expected to be
rather small for models with hierarchical neutrino masses, as discussed in \cite{Antusch:2007ib,CN}.

%%%%%%%%%%%%%%%%%%%%%%%%%%%%%%%

%%%%%%%%%%%%%%%%%%%%%%%%%%%%%%%

%%%%%%%%%%%%%%%%%%%%%%%%%%%%%%%

\section{Vacuum Alignment}
\label{Vacuum Alignment}
As is with flavour models of this type, the alignment of the vevs of the
flavon fields must be justified by minimising a flavon potential.  Thus, the
explicit directions of the vevs quoted in the previous section must be derived
from an explicit potential.  This is the goal of the present section.

As mentioned above, we have introduced the auxiliary flavon field
$\Phi^\nu_{\wt {3}}$ in order to align the other two neutrino flavons
which couple to the right-handed neutrinos. Furthermore, a set of ``driving
fields'' has to be added to generate the correct alignment of all 
flavon fields' vevs of the $\Delta(96)\times SU(5)$ model. These are listed  in
Table~\ref{driving} together with their transformation properties.
Driving fields are similar to flavons in that they are gauge singlets and
transform in a nontrivial way under $\Delta(96)\times U(1)\times Z_3$.
However, their difference becomes manifest when an additional $U(1)_R$
symmetry is introduced.\footnote{$U(1)_R$   is broken to $R$-parity when
  supersymmetry  breaking terms are included.} Under this symmetry, the
superspace variable $\theta$ is defined to have a $U(1)_R$ charge of $+1$.
Then, all chiral supermultiplets containing SM fermions also have a $U(1)_R$
charge of $+1$, supermultiplets containing Higgs fields and flavons are
neutral, while driving fields have a $U(1)_R$ charge of $+2$. A term allowed
in the superpotential itself carries a $U(1)_R$ charge of $+2$. This implies
that driving fields in the superpotential can only couple linearly to flavon
fields. With the further assumption that the driving fields develop positive soft
supersymmetric breaking scalar mass squared parameters at the symmetry
breaking scale, the driving fields do not develop vevs.  
Thus, it is only necessary to enforce that the $F$-terms of the driving fields
vanish identically. These so-called $F$-term conditions give rise to the
vacuum alignments.
In general the leading order terms of the flavon potential will be accompanied
by subleading operators. Imposing the $Z_3$ symmetry suppresses such
subleading terms, which couple the neutrino to the quark flavon fields 
(e.g. $X_1^u\Phi_2^u(\Phi_{\overline{3}}^d)^2\Phi^\nu_{\overline{3}^{\prime}}$ or 
$Y_1^d(\Phi_2^d)^2(\Phi_{\overline{3}}^d)^2\Phi^{\nu}_{\overline{3}^{\prime}}$), 
to a negligible level.  With these considerations in mind, we can begin to
create and minimise the flavon potential associated with the $\Delta(96)\times
SU(5)$ model.

\begin{table}[t]
\centering{\small{
\begin{tabular}{|c||c|c|c|c|c|c|c|c|c|c|}\hline
$\!$Field$\phantom{\Big|}\!\!\!$&$\!X_1^{\nu}\!$ & $\!X_2^{\nu}\!$ & $X_6^{\nu}$ 
& $X_1^d$& $Y_1^d$& $Z_1^d$
& $X_1^u$
& ${X}_1^{ud}$ &$X^{\nu d}_{1^{\prime}}$ & $X^{du}_2$  \\\hline
$\!\Delta(96)\phantom{\Big|}\!\!\!$&${\bf 1}$&${\bf 2}$&${\bf 6}$&${\bf 1}$&${\bf 1}$&${\bf 1}$&${\bf 1}$&${\bf 1}$&${\bf 1^{\prime}}$&${\bf 2}$\\\hline
$\!U(1)\phantom{\Big|}\!\!\!$&  $\!-6y\!$      &   $\!-4y\!$     &  $\!-2y-w\!$      &   $4y$  &  $\!-2z\!$&$\!x+3y+z\!$&$2x$& $\!2x+4y\!$ & $\!x+2y+2z-w\!$  &$\!2x-z\!$\\\hline
$\!Z_3\phantom{\Big|}\!\!\!$ &  $1$        & $\omega$    &$\omega$ & $1$ & $1$  & $1$ & $1$& $1$&  $\omega^2$   & $1$    \\\hline 
\end{tabular}}}
\caption{\label{driving} The driving fields required for the vacuum alignment
  of the $\Delta(96)\times SU(5)\times U(1)\times Z_3$ model.  All of these
  fields are singlets under $SU(5)$ and have a $U(1)_R$ charge of  $+2$.}
\end{table}

%%%%%%%%%%%%%%%%%%%%%%%%%%%%%%%%%

%%%%%%%%%%%%%%%%%%%%%%%%%%%%%%%%%

%%%%%%%%%%%%%%%%%%%%%%%%%%%%%%%%%

\subsection{Aligning 
$\boldsymbol{\langle\Phi^{\nu}_{\overline{3}^{\prime}}\rangle}$ and 
$\boldsymbol{\langle\Phi^{\nu}_{\widetilde{3}^{\prime}}\rangle}$}

The discussion of the alignment of the flavon fields in the $\Delta(96)\times
SU(5)$ model is begun in the neutrino sector with the alignment of the vevs of
the $\Phi^{\nu}_{\overline{3}^{\prime}}$ and
$\Phi^{\nu}_{\widetilde{3}^{\prime}}$ flavon fields.  Notice that both flavons
have the same charge under the $U(1)$ and $Z_3$ shaping symmetries of $2y$ and
$\omega$, respectively (see Table \ref{trans}).  
In order to align both neutrino flavon fields {\it separately}, 
the structure of the Kronecker products of $\Delta(96)$ is exploited.
We first discuss the flavon~$\Phi^\nu_{{\ol{3}'}}$. Its vev is aligned
using the auxiliary flavon field~$\Phi^{\nu}_{\widetilde{3}}$ as well as the
$\Delta(96)$ sextet driving field $X^{\nu}_6$. As the $U(1)$ charge of $X^{\nu}_6$
involves the parameter $w$, its pairing with the auxiliary neutrino flavon is
enforced. The product $X^{\nu}_6 \Phi^{\nu}_{\widetilde{3}}$ has a $U(1)$
charge of $-2y$ and could therefore couple to both neutrino flavon fields
$\Phi^{\nu}_{\overline{3}^{\prime}}$ and 
$\Phi^{\nu}_{\widetilde{3}^{\prime}}$. However,
the product ${\bf 6\otimes  \widetilde{3}=
3\oplus\overline{3}\oplus 3^{\prime}\oplus
  \overline{3}^{\prime}\oplus 6}$ (cf. Appendix~\ref{Appendix}) shows that
the $\Delta(96)$ Kronecker products only allow for the coupling
\be
X^{\nu}_6 \Phi^{\nu}_{\widetilde{3}} \Phi^\nu_{{\ol{3}'}} \ .
\ee
With the driving field being a sextet under $\Delta(96)$, we obtain six
$F$-term conditions which, using the CG coefficients of
Appendix~\ref{Appendix}, read

\begin{eqnarray}
\label{X6}
\left.
\begin{array}{c}
 \vev{\Phi^{\nu}_{\widetilde{3},1}}\vev{\Phi^{\nu}_{\overline{3}^{\prime},2}}+\omega ^2
\vev{ \Phi^{\nu}_{\widetilde{3},2}}\vev{ \Phi^{\nu}_{\overline{3}^{\prime},3}} +\omega
\vev{ \Phi^{\nu}_{\widetilde{3},3}}\vev{ \Phi^{\nu}_{\overline{3}^{\prime},1}} =0 \ , \\
\vev{ \Phi^{\nu}_{\widetilde{3},1}}\vev{\Phi^{\nu}_{\overline{3}^{\prime},1}}+\omega^2 
\vev{\Phi^{\nu}_{\widetilde{3},2} }\vev{\Phi^{\nu}_{\overline{3}^{\prime},2}}+\omega 
\vev{\Phi^{\nu}_{\widetilde{3},3}} \vev{\Phi^{\nu}_{\overline{3}^{\prime},3}} =0  \ ,  \\
\vev{\Phi^{\nu}_{\widetilde{3},1}} \vev{\Phi^{\nu}_{\overline{3}^{\prime},3}}+\omega^2 
\vev{\Phi^{\nu}_{\widetilde{3},2}} \vev{\Phi^{\nu}_{\overline{3}^{\prime},1}}+\omega 
\vev{\Phi^{\nu}_{\widetilde{3},3}} \vev{\Phi^{\nu}_{\overline{3}^{\prime},2}}  =0  \ , \\
\omega 
\vev{\Phi^{\nu}_{\widetilde{3},1} } \vev{\Phi^{\nu}_{\overline{3}^{\prime},2}}+\omega^2 
\vev{\Phi^{\nu}_{\widetilde{3},2} } \vev{\Phi^{\nu}_{\overline{3}^{\prime},3} } + 
\vev{\Phi^{\nu}_{\widetilde{3},3} } \vev{\Phi^{\nu}_{\overline{3}^{\prime},1}}=0  \ , \\
 \omega 
\vev{\Phi^{\nu}_{\widetilde{3},1}} \vev{\Phi^{\nu}_{\overline{3}^{\prime},1}}+\omega^2
\vev{\Phi^{\nu}_{\widetilde{3},2}} \vev{\Phi^{\nu}_{\overline{3}^{\prime},2}}   +
\vev{\Phi^{\nu}_{\widetilde{3},3}} \vev{\Phi^{\nu}_{\overline{3}^{\prime},3}} =0 \ , \\
  \omega 
\vev{\Phi^{\nu}_{\widetilde{3},1}}  \vev{\Phi^{\nu}_{\overline{3}^{\prime},3}}+\omega^2 
\vev{\Phi^{\nu}_{\widetilde{3},2}} \vev{\Phi^{\nu}_{\overline{3}^{\prime},1}}  +
\vev{\Phi^{\nu}_{\widetilde{3},3}} \vev{\Phi^{\nu}_{\overline{3}^{\prime},2}}=0 \ .
\end{array}\right.
\end{eqnarray}
Subtracting the fourth equation down from the first, the fifth from the
second, and the sixth from the third yields a set of three relations between
the components of $\vev{\Phi^{\nu}_{\widetilde{3}}}$ and $\vev{\Phi^{\nu}_{\overline{3}^{\prime}}}$: 
\begin{eqnarray}
\label{X6rel1}
\left.
\begin{array}{c}
\vev{\Phi^{\nu}_{\widetilde{3},1}}\vev{ \Phi^{\nu}_{\overline{3}^{\prime},2}}
=\vev{\Phi^{\nu}_{\widetilde{3},3}}\vev{ \Phi^{\nu}_{\overline{3}^{\prime},1}}\ ,\\
\vev{\Phi^{\nu}_{\widetilde{3},1}}\vev{ \Phi^{\nu}_{\overline{3}^{\prime},1}}
=\vev{\Phi^{\nu}_{\widetilde{3},3}}\vev{  \Phi^{\nu}_{\overline{3}^{\prime},3}}\ , \\
\vev{\Phi^{\nu}_{\widetilde{3},1}}\vev{ \Phi^{\nu}_{\overline{3}^{\prime},3}}
=\vev{\Phi^{\nu}_{\widetilde{3},3} }\vev{\Phi^{\nu}_{\overline{3}^{\prime},2}} \ .
\end{array}\right.
\end{eqnarray}
Similar logic can be used to further derive more relations by subtracting
$\omega^2$ multiplying the fourth, fifth and sixth lines of Eq.~(\ref{X6})
from the first, second and third lines, respectively: 
\begin{eqnarray}
\label{X6rel2}
\left.
\begin{array}{c}
\vev{\Phi^{\nu}_{\widetilde{3},2}}\vev{ \Phi^{\nu}_{\overline{3}^{\prime},3}}
=\vev{\Phi^{\nu}_{\widetilde{3},3}}\vev{ \Phi^{\nu}_{\overline{3}^{\prime},1}}\ ,\\
\vev{\Phi^{\nu}_{\widetilde{3},2}}\vev{ \Phi^{\nu}_{\overline{3}^{\prime},2}}
=\vev{\Phi^{\nu}_{\widetilde{3},3}}\vev{ \Phi^{\nu}_{\overline{3}^{\prime},3}}\ ,\\
\vev{\Phi^{\nu}_{\widetilde{3},2}}\vev{ \Phi^{\nu}_{\overline{3}^{\prime},1}}
=\vev{\Phi^{\nu}_{\widetilde{3},3}}\vev{ \Phi^{\nu}_{\overline{3}^{\prime},2}}\ .
\end{array}\right.
\end{eqnarray}
Multiplying all three relations in Eq.~(\ref{X6rel1}), and separately
all three relations in Eq.~(\ref{X6rel2}) one obtains 
\begin{eqnarray}
\label{X6rel3}
\begin{array}{c}
\vev{\Phi^{\nu}_{\widetilde{3},1}}^3
  \vev{\Phi^{\nu}_{\overline{3}^{\prime},1}}\vev{\Phi^{\nu}_{\overline{3}^{\prime},2}}\vev{\Phi^{\nu}_{\overline{3}^{\prime},3}}
=\vev{\Phi^{\nu}_{\widetilde{3},2}}^3 \vev{\Phi^{\nu}_{\overline{3}^{\prime},1}}\vev{\Phi^{\nu}_{\overline{3}^{\prime},2}}\vev{\Phi^{\nu}_{\overline{3}^{\prime},3}}
=\vev{\Phi^{\nu}_{\widetilde{3},3}}^3
  \vev{\Phi^{\nu}_{\overline{3}^{\prime},1}}\vev{\Phi^{\nu}_{\overline{3}^{\prime},2}}\vev{\Phi^{\nu}_{\overline{3}^{\prime},3}}\ .
\end{array}
\end{eqnarray}
If we now assume that none of the components of
$\langle\Phi^{\nu}_{\overline{3}^{\prime}}\rangle$ vanish,\footnote{If one
  component of either $\langle\Phi^{\nu}_{\overline{3}^{\prime}}\rangle$ or
  $\langle \Phi^{\nu}_{\widetilde{3}}\rangle$ vanishes, one can easily show by
  Eqs.~(\ref{X6rel1}) and (\ref{X6rel2}) that one flavon triplet 
will {\it not} develop any vev.} 
we end up with 
\begin{equation}
\vev{\Phi^{\nu}_{\widetilde{3},1}}^3=\vev{\Phi^{\nu}_{\widetilde{3},2}}^3=
\vev{\Phi^{\nu}_{\widetilde{3},3}}^3 \ ,
\end{equation}
which aligns the vev of the auxiliary flavon $ \Phi^{\nu}_{\widetilde{3}}$ as 
\begin{eqnarray}
\label{alignphi3t}
\langle \Phi^{\nu}_{\widetilde{3}}\rangle\propto \left(
\begin{array}{c}
\omega^{p_1}\\
\omega^{p_2}\\
\omega^{p_3}
\end{array}\right) ,
\end{eqnarray}
where $p_1$, $p_2$, $p_3$ can take the values of $0$, $1$, or $2$. It is
possible to show that Eqs.~\eqref{X6rel1} and \eqref{X6rel2} only allow solutions
with $p_1+p_2+p_3=0\mathrm{\,mod\,}3$. For the rest of this work, it is assumed that $p_1=p_2=p_3=0$.

The alignment of the neutrino flavon
$\langle\Phi^{\nu}_{\overline{3}^{\prime}}\rangle$ is obtained 
by using the various relations in Eqs.~(\ref{X6rel1}) and (\ref{X6rel2}) 
and the newly calculated alignment of $\langle\Phi^{\nu}_{\widetilde{3}}\rangle$.  
With the assumption that there are no relative phases on
$\langle\Phi^{\nu}_{\widetilde{3}}\rangle$ one immediately finds the desired
alignment of
\begin{eqnarray}
\label{alignvev3barp}
\langle \Phi^{\nu}_{\overline{3}^{\prime}}\rangle\propto \left(
\begin{array}{c}
1\\
1\\
1
\end{array}\right).
\end{eqnarray}
Now that $\langle\Phi^{\nu}_{\widetilde{3}}\rangle$ and
$\langle\Phi^{\nu}_{\overline{3}^{\prime}}\rangle$ are aligned, it is
necessary to align $\langle\Phi^{\nu}_{\widetilde{3}^{\prime}}\rangle$ to
finish the discussion of the alignment of the neutrino flavons.

The alignment of $\langle\Phi^{\nu}_{\widetilde{3}^{\prime}}\rangle$ can be
accomplished by taking advantage of the $\Delta(96)$ singlets contained in the products of
three neutrino flavons. Since, $\Phi^{\nu}_{\overline{3}^{\prime}}$ and
$\Phi^{\nu}_{\widetilde{3}^{\prime}}$ have the same $U(1)$ charge, any product
of three of them will couple to the same object. From the charges in Table
\ref{driving}, we see that this object is $X_1^{\nu}$, and the allowed flavon
potential terms are\footnote{The other possible product 
$\Phi^{\nu}_{\ol{3}^{\prime}}  \Phi^{\nu}_{\wt{3}^{\prime}} \Phi^{\nu}_{\wt{3}^{\prime}} $
  does not contain a $\Delta(96)$ singlet.}   
\begin{equation}
\frac{1}{M}\,X_1^{\nu}
\left[
g_0
\Phi^{\nu}_{\overline{3}^{\prime}}\Phi^{\nu}_{\overline{3}^{\prime}}\Phi^{\nu}_{\overline{3}^{\prime}} +
g_1
\Phi^{\nu}_{\overline{3}^{\prime}}\Phi^{\nu}_{\overline{3}^{\prime}}\Phi^{\nu}_{\widetilde{3}^{\prime}}+
g_2
\Phi^{\nu}_{\widetilde{3}^{\prime}}\Phi^{\nu}_{\widetilde{3}^{\prime}}\Phi^{\nu}_{\widetilde{3}^{\prime}}
  \right].
\end{equation}
With the alignment of Eq.~\eqref{alignvev3barp}, the first term, proportional
to the coupling constant $g_0$, vanishes identically and is therefore irrelevant for the
discussion of the $\Phi^{\nu}_{\widetilde{3}^{\prime}}$ vacuum.
The remaining terms give rise to the following $F$-term condition once the 
solution for the already aligned 
$\langle\Phi^{\nu}_{\overline{3}^{\prime}}\rangle$ is applied:
\begin{eqnarray}
\label{X1nualign3}
&&3 g_1 (\varphi ^{\nu}_{\overline{3}^{\prime}})^2 \left(\vev{\Phi
  ^{\nu}_{\widetilde{3}^{\prime},1}}+\vev{\Phi
  ^{\nu}_{\widetilde{3}^{\prime},2}}+\vev{\Phi^{\nu}_{\widetilde{3}^{\prime},3}}\right)
\nonumber\\
&&\hspace{6.9mm}-2 g_2 \left(\vev{\Phi
  ^{\nu}_{\widetilde{3}^{\prime},1}}^3+\vev{\Phi^{\nu}_{\widetilde{3}^{\prime},2}}^3+\vev{\Phi^{\nu}_{\widetilde{3}^{\prime},3}}^3-3\vev{\Phi
  ^{\nu}_{\widetilde{3}^{\prime},1}}
\vev{\Phi ^{\nu}_{\widetilde{3}^{\prime},2}} \vev{\Phi
  ^{\nu}_{\widetilde{3}^{\prime},3}} \right)=0\ . \hspace{10mm}
\end{eqnarray}
The next step would be to solve for the conditions on the alignment of
$\langle\Phi ^{\nu}_{\widetilde{3}^{\prime}}\rangle$, but notice that there
does not exist enough equations to obtain a unique solution for the alignment
of $\langle \Phi^{\nu}_{\widetilde{3}^{\prime}}\rangle$.  
This leads to the introduction of the last superpotential term required for
the alignment of the neutrino flavon fields' vevs:
\begin{eqnarray}
\label{X2nualign}
X_2^{\nu}\Phi^{\nu}_{\widetilde{3}^{\prime}}\Phi^{\nu}_{\widetilde{3}^{\prime}}
\ .
\end{eqnarray}
As the driving field $X_2^{\nu}$ is a $\Delta(96)$ doublet, which does not
couple to $\Phi^{\nu}_{\ol{3}^{\prime}}$, we obtain two simple $F$-term conditions
\begin{eqnarray}
\label{X2nualign2}
\begin{array}{c}
\vev{\Phi^{\nu} _{\widetilde{3}^{\prime},1}}^2+2 \vev{ \Phi^{\nu}
  _{\widetilde{3}^{\prime},2}}\vev{ \Phi^{\nu} _{\widetilde{3}^{\prime},3}}=0
\ , \\
\vev{\Phi^{\nu} _{\widetilde{3}^{\prime},3}}^2+2 \vev{\Phi^{\nu}
  _{\widetilde{3}^{\prime},1}}\vev{ \Phi^{\nu}
  _{\widetilde{3}^{\prime},2}}=0\ .
\end{array}
\end{eqnarray}
With the results of Eqs.~(\ref{X1nualign3}) and (\ref{X2nualign2}), there
exists enough constraints to align 
$\langle \Phi^{\nu}_{\widetilde{3}^{\prime}}\rangle$ properly.  
Looking at Eq.~(\ref{X2nualign2}), one finds two types of nontrivial solutions:
\begin{eqnarray}
\langle \Phi^{\nu} _{\widetilde{3}^{\prime}}\rangle\propto\left(
\begin{array}{c}
0\\
1\\
0
\end{array}\right)\text{ and }
\left(
\begin{array}{c}
1\\
-\frac{\omega^{p}}{2}\\
\omega^{2p}
\end{array}\right),\label{unique2}
\end{eqnarray}
where $p=0,1,2$.  
Notice that the solutions with $p=1,2$ are of the form given in
Eq.~(\ref{neutvevs}), hence these two solutions respect the desired low-energy
Klein symmetry. With this in mind, we restrict ourselves to this set when
solving Eq.~(\ref{X1nualign3}).  When this is done, four solutions are found:
\begin{eqnarray}
\langle \Phi^{\nu} _{\widetilde{3}^{\prime}}\rangle= \pm i \varphi^{\nu} _{\overline{3}^{\prime}}\omega^{2p}\sqrt{\frac{2g_1}{3g_2}}\left(
\begin{array}{c}
1\\
-\frac{\omega^p}{2}\\
\omega^{2p}\\
\end{array}\right),\label{unique4}
\end{eqnarray}
where  $p=1,2$. Comparing this result and Eq.~(\ref{alignvev3barp})
to Eq.~(\ref{neutvevs}), it is clear to see that 
$\langle \Phi^{\nu} _{\overline{3}^{\prime}}\rangle$ and
$\langle \Phi^{\nu} _{\widetilde{3}^{\prime}}\rangle$ are aligned in the
proper way to spontaneously break $\Delta(96)$ to the desired low-energy  Klein
symmetry in the neutrino sector, and are of approximately equal magnitude.  
With the flavons associated with the
neutrino sector properly aligned, the next task is to correctly align the
flavons associated with the charged leptons and quarks.  This endeavour will
begin with the alignment of the flavons which furnish doublet representations
of $\Delta(96)$.

%%%%%%%%%%%%%%%%%%%%%%%%%

%%%%%%%%%%%%%%%%%%%%%%%%%

%%%%%%%%%%%%%%%%%%%%%%%%%

\subsection{Aligning $\boldsymbol{\langle\Phi_2^d\rangle}$,
$\boldsymbol{\langle\Phi_2^u\rangle}$, and 
$\boldsymbol{\langle\bar{\Phi}_2^u\rangle}$}
The quest for the correct vacuum alignment for the flavon fields is continued
by considering the flavons which transform as doublets under
$\Delta(96)$, {\it i.e.} $\Phi_2^d$, $\Phi_2^u$, and $\bar{\Phi}_2^u$.  Considering
the charges in Tables~\ref{driving} and \ref{trans} for $Y_1^d$ and
$\Phi_2^d$, respectively, allows the flavon superpotential term 
\begin{equation}
Y_1^d\Phi_2^d\Phi_2^d =
2 Y_1^d\Phi_{2,1}^d\Phi_{2,2}^d \ .
\end{equation}
Clearly, the resulting $F$-term condition has two solutions
\begin{eqnarray}
\langle \Phi_2^d\rangle \propto \left(
\begin{array}{c}
1\\
0
\end{array}\right) \text{ or}
\left(
\begin{array}{c}
0\\
1
\end{array}\right),\label{unique1}
\end{eqnarray}
of which we choose the alignment consistent with Eq.~(\ref{vevphi2d}), \textit{i.e.}
the alignment in which $\langle\Phi_{2,2}^d\rangle$ vanishes.  

Turning to the $X^{du}_2$ driving field, we find the flavon potential coupling
\begin{equation}
X^{du}_2\Phi_2^d\Phi_2^u
=X^{du}_{2,1}\Phi_{2,1}^d\Phi_{2,1}^u+X^{du}_{2,2}\Phi_{2,2}^d\Phi_{2,2}^u
\ .
\end{equation}
With the already aligned $\vev{\Phi^d_2}$, the $F$-term condition of
$X^{du}_{2,2}$ is automatically satisfied, while the condition arising from the
$F$-term of $X^{du}_{2,1}$ enforces the alignment
\begin{eqnarray}
\label{vevphi2u}
\langle \Phi_2^u \rangle\propto \left(
\begin{array}{c}
0\\
1
\end{array}\right).
\end{eqnarray}

Finally, we arrive to the last doublet to align, $\vev{\bar{\Phi}_2^u}$.  From the
charges given in Tables \ref{driving} and \ref{trans}, it is seen that
$\bar{\Phi}_2^u$ couples to $\Phi_2^u$ and $X_1^u$ as 
\begin{equation}
X_1^u\Phi_2^u\bar{\Phi}_2^u=X_1^u(\Phi^u_{2,1}\bar{\Phi}^u_{2,2}+\Phi^u_{2,2}\bar{\Phi}^u_{2,1}).
\end{equation}
Inserting the vacuum alignment for $\vev{\Phi_2^u}$ given in Eq.~(\ref{vevphi2u}) immediately implies that 
\begin{eqnarray}
\label{phitilde2u}
\langle \bar{\Phi}_2^u\rangle \propto\left(
\begin{array}{c}
0\\
1
\end{array}\right).
\end{eqnarray}
With the derivation of this last result, the vevs of all flavons transforming
as a doublet of $\Delta(96)$ have been aligned accordingly.  Thus, the
remaining task is to align the vevs of $\Phi_{\overline{3}}^d$ and
$\bar{\Phi}_{\overline{3}}^d$.

%%%%%%%%%%%%%%%%%%%%%%%%%

%%%%%%%%%%%%%%%%%%%%%%%%%

%%%%%%%%%%%%%%%%%%%%%%%%%

\subsection{Aligning 
$\boldsymbol{\langle\Phi_{\overline{3}}^d\rangle}$ and 
$\boldsymbol{\langle\bar{\Phi}_{\overline{3}}^d\rangle}$}

The last set of flavon fields to align is that set of flavons contributing to
the down quark and charged lepton masses and mixings, $\Phi_{\overline{3}}^d$
and $\bar{\Phi}_{\overline{3}}^d$.  This final task is begun by
considering~$\langle\Phi_{\overline{3}}^d\rangle$.  From Table \ref{driving},
it can be seen that the leading flavon potential terms which contribute to
the alignment are 
\be
\frac{1}{M^2}X^d_1\Phi_{\overline{3}}^d\Phi_{\overline{3}}^d\Phi_{\overline{3}}^d\Phi_{\overline{3}}^d
~+~
\frac{1}{M^3}{X}^{ud}_1\Phi_2^u\Phi_{\overline{3}}^d\Phi_{\overline{3}}^d\Phi_{\overline{3}}^d\Phi_{\overline{3}}^d
\ ,\label{ud1alig}
\ee
which lead to the $F$-term conditions
\be
\label{X1d}
\left(\vev{\Phi ^d_{\overline{3},2}}^2+2 \vev{\Phi ^d_{\bar{3},1}}\vev{ \Phi
  ^d_{\overline{3},3}}\right)^2
+2 \left(\vev{\Phi ^d_{\overline{3},1}}^2+2 \vev{\Phi ^d_{\overline{3},2}}
\vev{\Phi^d_{\overline{3},3}}\right) \!
\left(\vev{\Phi ^d_{\overline{3},3}}^2+2 \vev{\Phi ^d{}_{\overline{3},1}}\vev{
  \Phi ^d_{\overline{3},2}}\right) = 0 \ ,
\ee
and
\be
\label{Xbar1d}
\left(\vev{\Phi ^d_{\overline{3},1}}^2+2 \vev{\Phi ^d_{\bar{3},2}}\vev{ \Phi
^d_{\overline{3},3}}\right)^2
+2 \left(\vev{\Phi ^d_{\overline{3},2}}^2+2 \vev{\Phi
^d_{\overline{3},1}} \vev{\Phi^d_{\overline{3},3}}\right)\!
 \left(\vev{\Phi ^d_{\overline{3},3}}^2+2
\vev{\Phi ^d{}_{\overline{3},1}} \vev{\Phi ^d_{\overline{3},2}}\right) = 0 \ .
\ee
Note that we have used the already calculated alignment of
$\langle\Phi_2^u\rangle$ in order to derive Eq.~(\ref{Xbar1d}) from the second
term of Eq.~\eqref{ud1alig}. 
These two $F$-term conditions allow for the following sixteen vacuum alignments:
\begin{eqnarray}
\label{phi3barvev2}
\langle\Phi^d_{\overline{3}}\rangle\propto\left(
\begin{array}{c}
0\\
0\\
1
\end{array}\right),
\left(\begin{array}{c}
\omega^{q_1}\\
1\\
-\frac{\omega^{2q_1}}{2}
\end{array}\right),
\left(\begin{array}{c}
\omega^{ q_2}\\
1\\
(-\frac{1}{2}\pm \frac{3i}{2}) \omega^{2 q_2}
\end{array}\right),
\left(\begin{array}{c}
(-2\pm\sqrt{3})\omega^{q_3}\\
1\\
(-\frac{1}{2}\pm\frac{\sqrt{3}}{2})\omega^{2 q_3}
\end{array}\right),
\end{eqnarray}
where $q_1$, $q_2$, and $q_3$ take the values of $0$, $1$, or $2$.  We select
the first solution in Eq.~(\ref{phi3barvev2}), as it is consistent with the
assumed alignment of Eq.~(\ref{vevphi2d}).  Now that
$\langle\Phi^d_{\overline{3}}\rangle$ has been aligned there exists only one
more flavon vev to align, $\langle\bar{\Phi}_{\overline{3}}^d\rangle$.

The alignment of $\langle\bar{\Phi}_{\overline{3}}^d\rangle$ is derived from 
the driving fields $X^{\nu d}_{1'}$ and $Z^d_1$. As can be seen from
Table~\ref{driving}, the relevant flavon potential terms read
\be
\frac{1}{M}X^{\nu d}_{1^{\prime}}\Phi^{\nu}_{\widetilde{3}}
\Phi^d_{\overline{3}} \bar{\Phi}^d_{\overline{3}}
+
\frac{1}{M^2}Z_1
^d\Phi_2^d\Phi_{\overline{3}}^d\Phi_{\overline{3}}^d\bar{\Phi}^d_{\overline{3}}
\ .
\ee
Inserting the already determined flavon alignments, we arrive at the $F$-term
conditions
\bea
\vev{\bar{\Phi }^d_{\overline{3},1}}+\vev{\bar{\Phi
  }^d_{\overline{3},2}}+\vev{\bar{\Phi }^d_{\overline{3},3}} \!\!&=&\!\! 0
\ ,\\
\vev{\bar{\Phi }^d_{\overline{3},1}}\!\!&=&\!\! 0 \ .
\eea
Therefore, the alignment of $\langle\bar{\Phi}_{\overline{3}}^d\rangle$ is
fixed uniquely as
\begin{eqnarray}
\langle\bar{\Phi}^d_{\overline{3}}\rangle\propto\left(
\begin{array}{c}
0\\
1\\
-1
\end{array}\right) ,
\end{eqnarray}
which is in agreement with the desired alignment stated in
Eq.~(\ref{vevphi2d}).

Before concluding this section, we briefly comment on the (non-)uniqueness of
the achieved set of alignment vectors. The fact that the individual $F$-term
conditions often yield multiple solutions, as seen e.g. in Eqs.~(\ref{unique1}) and
(\ref{phi3barvev2}), is a reflection of the symmetry properties of the
equations: finding one solution, we can easily obtain other solutions by
applying appropriate $\Delta(96)$ transformations. We have checked in each
case that all obtained solutions are indeed related to each other in such a
way. 

Indeed, if we consider the complete flavon potential involving all flavon fields, it is
generally true that a potential which gives the desired {\it set} of
alignments, will also give any $\Delta(96)$ transformed set of alignments,
{\it i.e.} a set of alignments where all vevs are transformed by the same
arbitrary $\Delta(96)$ group element. Being symmetry transformations, all
$\Delta(96)$ transformed sets of alignments are physically
equivalent. However, some sets of alignments are more convenient than others
when it comes to explicitly constructing a model of flavour, as done in
Section~\ref{Model}. Therefore we are interested in generating only our desired
alignments. Now, in the case of the flavon potential presented here, we face
the problem that not all sets of alignments which we can generate are related
by $\Delta(96)$ transformations. This means Nature could choose to fall into a
vacuum which is not related to the desired set of alignments corresponding to
the BT vacuum. In order to understand how much freedom Nature has, it is
useful to consider the alignments of the various flavons in turn. 

Starting with Eq.~(\ref{unique1}) it is clear that both solutions are related
by a $U$ transformation, {\it i.e.} we can choose the desired alignment of 
$\vev{\Phi^d_2}$ without loss of generality. The other two doublet flavons 
$\vev{\Phi^u_2}$ and $\vev{\bar \Phi^u_2}$ are then aligned without any
ambiguities. 
Continuing with Eq.~(\ref{phi3barvev2}), we find sixteen
solutions. One can show explicitly that these are related to the desired
$\vev{\Phi^d_{\ol 3}}$ alignment by $\Delta(96)$ transformations involving an
even number of $U$ factors (and a collection of $S$ and $T$
factors). Such a group transformation can be used to choose the desired
$\vev{\Phi^d_{\ol 3}}$ alignment, but at the same time, the same
transformation must also be applied to the already aligned doublet flavon
vevs. Involving an even number of $U$ factors, one quickly finds that the form
of the doublet flavons remains unaltered by such a  transformation (except for
possible factors of $\omega$). 
Next, we consider the alignment of $\vev{\Phi^\nu_{\wt 3}}$ whose general
solution is given in Eq.~\eqref{alignphi3t}. Using a $T$ transformation we can
bring this into the standard $(1,1,1)^T$ form. Notice that such a $T$
transformation does not change the above flavon alignments but only multiplies 
the vevs by possible factors of $\omega$. 
With the choices made so far, the alignments of $\vev{\Phi^\nu_{\ol  3'}}$ and
$\vev{\bar\Phi^d_{\ol 3}}$ are determined uniquely. 
The ensuing set of seven alignments is obtained without loss of
generality, since it is related to all other sets of the same seven alignments
derived from the flavon potential terms by $\Delta(96)$ transformations. 
In the next and final step we have to consider the alignment of 
$\vev{\Phi^\nu_{\wt 3'}}$, cf. Eq.~\eqref{unique2}. 
The solution on the left is related to the three solutions on the right by
$\Delta(96)$ transformations, namely $T^k U$, where $k=0,1,2$. 
Applying such a transformation to all the other seven flavon alignments
changes at least one of the them, and the desired set of
alignments is not obtained any longer. Therefore, it is not possible to
choose the desired alignment for~$\vev{\Phi^\nu_{\wt 3'}}$ without loss of
generality. In fact, the four solutions for the alignment
of~$\vev{\Phi^\nu_{\wt 3'}}$ generated four physically different sets of
alignments which are not related by $\Delta(96)$. Two of them
correspond to BT vacua, while the other two predict unphysical mixing angles.
In order for our $\Delta(96)$ model to be viable, it is therefore necessary
for Nature to fall into one of the two BT vacua. As the flavon potential
features only four possible vacua, this is a very mild assumption.

%%%%%%%%%%%%%%%%%%%%%%%%%%%%%%%%%%%

%%%%%%%%%%%%%%%%%%%%%%%%%%%%%%%%%%%

%%%%%%%%%%%%%%%%%%%%%%%%%%%%%%%%%%%

\section{The Subleading Order}
\label{NLO}
In this section, we briefly discuss the next to leading order (NLO) contributions to the superpotential
arising from choosing the $U(1)$ charge parameters ($w$, $x$, $y$, and $z$), in Tables \ref{trans} and \ref{driving}, as unrelated.  Note that for any given set of integer choices for $w$, $x$, $y$, and $z$ we expect additional model-dependent NLO terms not discussed in this section.  Here we only present the inevitable NLO terms resulting from such a charge assignment.  By making this assumption, we can determine all minimal NLO terms allowed by the $U(1)$ and $Z_3$
symmetries. To start with, we ignore the $\Delta(96)$ transformation
properties and find that 
there are two $U(1)$ neutral factors
\be
A=\frac{1}{M} \bar\Phi^u_2 \ , \qquad 
B=\frac{1}{M^3}(\Phi^d_{\ol 3})^2 \Phi^\nu_{{\ol 3'},{\wt 3'}} \ ,
\ee
where we have introduced a unified notation for the two neutrino type flavons fields
$\Phi^\nu_{{\ol 3'}}$ and $\Phi^\nu_{{\wt 3'}}$.
$A$ is of order $\lambda^4$, and $B$ of order $\lambda^8$ since we
assume that the neutrino flavon vevs are of order $\lambda^4 M$. 
$A$ is neutral under $Z_3$ while $B$ carries charge $\omega$.
Provided the $\Delta(96)$ symmetry is  satisfied one can multiply
each LO term that we give by either $A$ or $B^3$. Clearly the latter is
completely irrelevant as it is suppressed by $\lambda^{24}$. The former can
only give corrections of order $\lambda^4$ or smaller (in case one power of
$A$ is not sufficient to generate a $\Delta(96)$ invariant term).
Following the above logic and 
also demanding invariance under both $\Delta(96)$ and $U(1)\times Z_3$ symmetry groups, it is found that the leading
NLO corrections to the Yukawa superpotential can arise from operators of the form:
\bea
FT_3H_{\overline{5}}\frac{1}{M}\Phi^d_{\overline{3}}A, \ \ \ \ 
FTH_{\overline{45}}\frac{1}{M^2}\bar{\Phi}^d_{\overline{3}}\Phi^d_2 A, \ \ \ \ 
FTH_{\overline{5}}\frac{1}{M^3}\bar{\Phi}_{\overline{3}}^d(\Phi_2^d)^2  A,  \\\nonumber
TTH_5\frac{1}{M^2}\Phi_2^u\bar{\Phi}_2^u A, \ \ \ \ 
T_3T_3H_5A^2, \ \ \ \
FNH_5 A, \ \ \ \
NN\Phi^{\nu}_{{\ol 3'},{\wt 3'}}A. ~\,
\eea
In order to explain the abundance of $A$ couplings in the above set of
operators, recall that having a $\bar{\Phi}_2^u\sim A$ flavon neutral under
the $U(1)$ symmetry is essential for generating the up quark mass suppression
in this model [cf. Eq. (\ref{wu})]. Since $\bar{\Phi}_2^u\approx\lambda^4 M$,
it is seen that these NLO corrections are relatively small compared to their
LO counterparts and can be neglected, {\it i.e.} the ratio of NLO to LO Yukawa
couplings for each contributing term to the Yukawa matrix is equal to or
smaller than $\lambda^4$. As an example, consider
$FT_3H_{\overline{5}}\frac{1}{M}\Phi^d_{\overline{3}}A$.  Regardless of the 
contraction, this term can only contribute to the third row (column) of
$M_{d}$ ($M_e$) at an order of $\lambda^6$, as such also filling in the
previously existing zeroes. Comparing this result to  the schematic
suppressions of the mass matrices given in Eq. (\ref{masssup}), it is found
that such NLO effects are negligible. Other than those terms discussed above,
we also find a couple new terms not derived from the LO terms. Demanding
invariance under the $U(1)$ and $Z_3$ symmetries they are as follows: 
\be
TT_3H_{5} \frac{1}{M^5} 
\Phi^d_{\ol 3} 
\bar\Phi^d_{\ol 3} 
(\Phi^d_2)^2
 \Phi^\nu_{{\ol 3'},{\wt 3'}}  B^2 \ ,\qquad
TTH_{5} \frac{1}{M^7} 
(\bar\Phi^d_{\ol 3} )^2
(\Phi^d_2)^4
 \Phi^\nu_{{\ol 3'},{\wt 3'}}  B^2 \ .
\ee
Due to the factor $B^2$, these terms are certainly very suppressed and can be
ignored.

A similar result is obtained when calculating the leading NLO corrections to the flavon superpotential.  We find that all flavon superpotential terms receive an NLO correction related to the coupling of $A$ to the existing LO term. In addition, we find several new operators:
\bea
&&X^u_1 \frac{1}{M^4}
(\bar\Phi^d_{\ol 3} )^2
(\Phi^d_2)^4 \ B^3, \\
&&X^{ud}_1 \frac{1}{M^6}
(\Phi^d_{\ol 3} )^2
(\bar\Phi^d_{\ol 3} )^2
(\Phi^d_2)^4 \ ,\label{fourthterm1}\\
&&X^{du}_2 \frac{1}{M^6}
\Phi^d_{\ol 3} 
(\bar\Phi^d_{\ol 3} )^2
(\Phi^d_2)^5 \ .
\eea
Notice that, in the above, a factor of $B^3$ was added to the first operator to enforce $\Delta(96)$ invariance.  As a result of this, this operator is clearly heavily suppressed.  Yet, the second term is of order $\lambda^{14}$ which is to be compared to the LO
term involving that driving field which would be of order $\lambda^{12}$. Thus,
here there exists a more significant NLO term, which is related to the
alignment of $\Phi^d_{\ol 3}$.
Finally the third term is of order $\lambda^{13}$ which needs to be compared to the LO
term involving that driving field, a term which is of order $\lambda^5$. As
the relative importance of the NLO term is $\frac{\lambda^{13}}{\lambda^5} =
\lambda^8$, it is negligible.

In summary, the majority of the NLO to LO correction ratios are of order $\lambda^4$ (or smaller) with the exception of the correction to the flavon potential involving the driving field $X^{ud}_1$ [cf. Eq. (\ref{fourthterm1})].  As mentioned, this term contributes an $F$-term of order $\lambda^{14}$. 
Comparing to the LO term involving $X_1^{ud}$
in Eq.~\eqref{ud1alig}
it is seen that the ratio of NLO to LO $F$-terms is suppressed by $\lambda^2$.
In principle this term could lead to significant corrections to vacuum alignment of the d-type flavons
in Eq.~(\ref{vevphi2d}), possibly filling in the zeros at order $\lambda^2$,
which would have phenomenological effects.
However, in practice, by plugging in the leading order alignment
into Eq.~\eqref{fourthterm1}, such a correction vanishes.
In other words the NLO correction arising from the operator in Eq.~\eqref{fourthterm1}
is consistent with an alignment proportional to the LO result. As a consequence,
there will not be any distortion of the vacuum alignment induced by this term.

%%%%%%%%%%%%%%%%%%%%%%%%%%%%%%%%%%%

%%%%%%%%%%%%%%%%%%%%%%%%%%%%%%%%%%%

%%%%%%%%%%%%%%%%%%%%%%%%%%%%%%%%%%%

\section{Conclusion}
\label{Conclusion}

Recent results from the Daya Bay and RENO Collaborations have shown that the
reactor mixing angle is non-zero and quite sizeable.  This result
presents a new challenge for the existing paradigms of discrete flavour
symmetries which attempt to describe all quark and lepton masses and mixing
parameters. In the new era of a non-zero reactor angle, 
$\Delta(96)$ is a very promising candidate family symmetry since it is 
capable of predicting all the lepton mixing angles, 
including a quite sizeable reactor angle. However, the resulting simple
Bi-trimaximal mixing pattern obtained from one embedding of the Klein symmetry
(the most promising one)
gives mixing angles outside the global fit ranges, and hence simple models of
leptons based on  $\Delta(96)$ are not viable. This motivates going beyond the
simple models of leptons proposed so far, and in particular to GUT models
where modest charged lepton corrections can correct the BT predictions,
bringing them into agreement with the recent global fits. 

In this paper we have proposed a SUSY GUT of Flavour based upon an $SU(5)$ gauge group,
together with a $\Delta(96)$ flavour symmetry. In particular, we
considered that the $\bf{\overline{5}}$-plets of $SU(5)$ transform as a ${\bf 3}$
under the $\Delta(96)$ flavour group. We made use of the ${\bf 2}$ representation
of $\Delta(96)$ by assigning the two ${\bf 10}$-plets
of $SU(5)$ corresponding to the lightest two families to transform as a ${\bf 2}$,
with the ${\bf 10}$
associated with the third family transforming in the trivial 
${\bf 1}$ under $\Delta(96)$.  In the Higgs sector, a ${\bf \overline{45}}$
field was added to obtain the GJ mass relations.
Right-handed neutrino fields, $N$, transforming as singlets under $SU(5)$ and
${\bf\overline{3}}$ under $\Delta(96)$ were also added in order to generate
neutrino masses and mixings.
The family symmetry is broken by a set of eight gauge singlet flavon fields, leading to  
phenomenologically viable fermion mass matrices. Finally, an additional $U(1)$
and $Z_3$ symmetry was employed to prevent the proliferation of unwanted
terms in the superpotential.

%%%%%%%%%%%%%%%%%%%%%%%%%%%%%%%%%%%%%%%%%%%%%%%%%%%%%%%%%%%%%%%%
The model clearly contains a large number of flavon fields which couple to the quarks
and charged leptons through effective non-renormalisable operators. This is a common
feature of models aiming to explain the flavour structure of the Standard
Model fermions in the framework of a supersymmetric GUT (see
Refs.~\cite{S4GUT,Hagedorn:2012ut,SUSYGUTs} for other examples of such
models). It is instructive to construct such models as they showcase
successful concepts applied in advancing our understanding of flavour, but also
their current limitations leading to new ideas. This provides the motivation for 
constructing such models. The assumptions concerning the present model are summarised in the 
following paragraph.

The flavons of the $\Delta(96)$ model are classified according to their LO
Yukawa couplings: we introduced two flavons for the up-type quarks, three
flavons for the down-type quarks and charged leptons, and finally two flavons
for the neutrinos. The auxiliary third neutrino flavon of Table~\ref{trans} does not
enter the Yukawa sector, but played an important role in aligning the other
neutrino flavons. The effective non-renormalisable terms are suppressed by powers of a
suppression scale $M$ which, for simplicity, we assumed to be common to all
operators. In some cases, c.f. Eq. (\ref{wd}), specific $\Delta(96)$
contractions of flavon and quark/charged lepton fields are needed 
in order to yield phenomenologically viable predictions. Thus, we postulated
that above the scale $M$ a set of mediator fields exists which selects out
the desired contractions.\footnote{Explicit examples of mediator field
arguments can be found in Refs.~\cite{S4GUT,Hagedorn:2012ut,UVcomplete}.}   
In order to construct the mass
matrices associated with the Yukawa operators given in
Eqs.~(\ref{wu})-(\ref{wnu}), a set of vacuum alignments
was justified in Section~\ref{Vacuum Alignment}.  
The model does not predict the scale of the flavon vevs, which we
simply assumed to generate the observed fermion mass hierarchies. However, we
emphasise that the scales of the flavons in Eq. (\ref{wnu}) were shown to
be related to each other when considering the vacuum alignment in Section
\ref{Vacuum Alignment}.  
The above assumptions characterise the considered $\Delta(96)\times SU(5)$ theory of flavour. 
Disregarding the dimensionless order one coupling constants in the Yukawa and flavon potential,
we have to fix only six free input parameters. These correspond to the vevs
of the seven flavon fields coupling to the fermions minus one condition which
relates the two flavons of the neutrino sector.

Having recapitulated our motivations and assumptions, the leading
order Yukawa terms associated with the up, down, charged lepton, and neutrino
sectors, after the $\Delta(96)$ flavour symmetry and electroweak symmetry breaking, 
lead to a corresponding set of mass matrices discussed in Section~\ref{Model}.
The model describes the strong mass hierarchy among the up-type quarks,
$m_u: m_c: m_t\approx \lambda^8: \lambda^4 : 1$ with no mixing in the up
sector at leading order. It also reproduces the weaker down-type quark mass hierarchy
$m_d: m_s: m_b\approx \lambda^4: \lambda^2: 1$,
with quark mixing angles satisfying the GST relation
(\textit{i.e.} $\theta^q_{12}\approx\theta^d_{12}\approx\sqrt{m_d/m_s}$).
The GUT-scale GJ relations $m_e/m_d = 1/3$, $m_{\mu}/m_s=3$, $m_{\tau}/m_b= 1$
also emerge. However we emphasise that the only relevant aspect of the GJ relations
for predicting the PMNS angles is the left-handed charged lepton
mixing angle prediction $\theta^e_{12}\approx \lambda / 3$,
and this could equally well emerge from alternatives to GJ, for example those that predict $m_{\mu}/m_s=9/2$.

Applying the charged lepton mixing correction $\theta^e_{12}\approx \lambda / 3$
with zero phase to the approximate leading order
  BT values for the atmospheric, solar, and reactor mixing angles, yields
$\theta_{23}\approx 36.9^{\circ}$, $\theta_{12}\approx 32.7^{\circ}$ and
$\theta_{13}\approx 9.6^{\circ}$, respectively, at the GUT scale, in good
  agreement with recent global fits and leading to the prediction of a zero
  Dirac CP phase $\delta \approx 0$. In general, including charged lepton
  corrections arising from any GUT model involving $\theta^e_{12}\approx
  \lambda / 3$, would correct the $\Delta(96)$ predictions 
almost perfectly, providing that the charged lepton corrections  
carry a zero phase. This latter feature must eventually must be explained within
a more complete theory beyond the present model, where all phases are predicted,
for example along the lines of the models proposed in \cite{Antusch:2011sx}.

\section*{Acknowledgements}
The authors acknowledge partial support from the EU ITN grants UNILHC
  237920 and INVISIBLES 289442. The work of SFK and AJS is supported by the STFC
Consolidated  ST/J000396/1 grant. SFK and CL thank the Galileo Galilei
Institute for Theoretical Physics for hospitality.

%%%%%%%%%%%%%%%%%%%%%%%%%

%%%%%%%%%%%%%%%%%%%%%%%%%

%%%%%%%%%%%%%%%%%%%%%%%%%

\appendix
\section{Appendix: The Group Theory of ${\boldsymbol{\Delta(96)}}$}
\label{Appendix}
\subsection{The Structure of ${\boldsymbol{\Delta(96)}}$}

The group $\Delta(96)$ is a non-Abelian discrete subgroup of $SU(3)$ of order
96.  In fact, it is the $\Delta(6n^2)$ group with $n=4$ (see
Ref.~\cite{Escobar:2008vc} for a detailed discussion of $\Delta(6n^2)$
groups).  Thus, 
\begin{equation}
\Delta(96)\cong (Z_4\times Z_4)\rtimes S_3.
\end{equation}
Furthermore, it has 10 conjugacy classes.  They are $I$ (the trivial conjugacy class),
$3C_4$, $3C_2$, $3C_4^{\prime}$, $6 C_4^{\prime\prime}$, $32C_3$,
$12C_4^{\prime\prime\prime}$, $12C_8$, $12C_2^{\prime}$, and
$12C_8^{\prime}$\cite{Delta(96)orig1,Escobar:2008vc}.  In this list of
conjugacy classes, we have adopted Schoenflies notation in which the number
in front of a given conjugacy class, $C_n$, is the number of elements belonging to it and the subscript ``$n$''
denotes the order of the elements contained in it.  As a
result of these conjugacy classes and the theorems that prove that the number of
irreducible representations is equal to the number of conjugacy classes and the sum of the squares of the dimensions of the
irreducible representations is equal to the order of the group, it
is easy to see that $\Delta(96)$ has two singlet irreducible representations (${\bf 1}$
and ${\bf 1^{\prime}}$), one doublet irreducible representation (${\bf 2}$), six triplet irreducible representations (${\bf 3}$, ${\bf \widetilde{3}}$, ${\bf \overline{3}}$, ${\bf
  3^{\prime}}$, ${\bf \widetilde{3}^{\prime}}$, ${\bf
  \overline{3}^{\prime}}$), and one sextet
(${\bf 6}$) irreducible representation because 
\begin{equation}
1+3+3+3+6+32+12+12+12+12
=96=1^2+1^2+2^2+3^2+3^2+3^2+3^2+3^2+3^2+6^2.
\end{equation}
With these irreducible representations and conjugacy classes, it is possible
to write down the character table for $\Delta(96)$ by applying the logic in
Ref.~\cite{Escobar:2008vc}, see Table \ref{Character Table}.
\begin{table}[t]
\centering
\begin{tabular}{|c|c|c|c|c|c|c|c|c|c|c|}
\hline
 $\Delta(96)$&$\bf{1}$ & $\bf{1^{\prime}}$ & $\bf{2}$ &$\bf{3}$ &$\bf{ \widetilde{3}}$&$\bf{\overline{3}}$ & $\bf{3^{\prime}}$&$\bf{ \widetilde{3}^{\prime}}$ & $\bf{\overline{3}^{\prime}}$ &$ \bf{6} $\\\hline
$\mathcal{I}$& $1$ & $1$ & $2$ & $3$ & $3$ & $3$ & $3$ & $3$ & $3$ & $6$ \\\hline
$3C_4$& $ 1$ & $1$ & $2$ & $-1+2 i$ & $-1$ & $-1-2 i$ & $-1+2 i$ & $-1$ &$ -1-2 i$ & $2$ \\\hline
$3C_2$&  $1$ & $1$ & $2$ & $-1$ & $3$ & $-1$ & $-1$ & $3$ & $-1$ & $-2$ \\\hline
$3C_4^{\prime}$&  $1$ & $1$ & $2$ & $-1-2 i$ & $-1$ & $-1+2 i$ & $-1-2 i$ & $-1$ & $-1+2 i$ & $2$ \\\hline
 $6 C_4^{\prime\prime}$ &$1$ & $1$ & $2$ & $1$ & $-1$ & $1$ & $1$ & $-1$ & $1$ & $-2$ \\\hline
 $32C_3$& $1$ & $1$ & $-1$ & $0$ & $0$ & $0$ & $0$ & $0$ & $0$ & $0$ \\\hline
$12C_2^{\prime}$&  $1$ & $-1$ & $0$ & $-1$ & $-1$ & $-1$ & $1$ & $1$ & $1$ & $0$ \\\hline
 $12C_8$ & $1$ & $-1$ & $0$ & $i$ & $1$ & $-i$ & $-i$ & $-1$ & $i$ & $0$ \\\hline
 $12C_4^{\prime\prime\prime}$ & $1$ & $-1$ & $0$ & $1$ & $-1$ & $1$ & $-1$ & $1$ & $-1$ & $0$ \\\hline
 $12C_8^{\prime}$ & $1$ & $-1$ & $0$ & $-i$ & $1$ & $i$ & $i$ & $-1$ & $-i$ & $0$\\\hline
\end{tabular}
\caption{The Character Table of $\Delta(96)$}
\label{Character Table}
\end{table}

With the character table, it is easy to calculate the Kronecker products of
$\Delta(96)$. See Table \ref{table:kronecker} for a complete listing of the
Kronecker products of $\Delta(96)$.  These Kronecker products will aid in the
calculation of CG coefficients for the decomposition of the product
representations of $\Delta(96)$.  However, this cannot be done until an
explicit representation (and presentation) is chosen for each of
$\Delta(96)$'s irreducible representations.

\begin{table}[t]
\centering
\begin{tabular}{|c|}
\hline
$\bf{1}\otimes x=x$ with $\bf{x}$ any $\Delta(96)$ irrep\\
$\bf{1^{\prime}}\otimes \bf{1^{\prime}}=\bf{1}$\\
$\bf{1^{\prime}\otimes 2=2}$\\
$\bf{1^{\prime}\otimes r=r^{\prime}}$ when $\bf{r=3}$, $\bf{\widetilde{3}}$, or $\bf{\overline{3}}$\\
$\bf{1^{\prime}\otimes r^{\prime}=r}$ when $\bf{r=3}$, $\bf{\widetilde{3}}$, or $\bf{\overline{3}}$\\
$\bf{1^{\prime}\otimes 6=6}$\\
$\bf{2\otimes 2=1\oplus 1^{\prime}\oplus 2}$\\
%$2\otimes r=r\oplus r^{\prime}$ when $r=3$, $\widetilde{3}$, or $\overline{3}$\\
$\bf{2}$$\otimes\bf{r}$$^{m}=\bf{r\oplus r^{\prime}}$ when $\bf{r=3}$, $\bf{\widetilde{3}}$, or $\bf{\overline{3}}$\\
$\bf{2\otimes 6=6\oplus 6}$\\
$\bf{3}$$^m\otimes \bf{3}$$^{n}=\bf{\widetilde{3}}$$^{p}\oplus \bf{\overline{3}^{\prime}\oplus \overline{3}}$\\
$\bf{3}$$^m\otimes \bf{\widetilde{3}}$$^{n}= \bf{\overline{3}}$$^{p}\oplus \bf{6}$\\
$\bf{3}$$^m\otimes \bf{\overline{3}}$$^{n}=\bf{1}$$^{q}\bf{\oplus 2\oplus 6}$\\
$\bf{\widetilde{3}}$$^m\otimes \bf{\widetilde{3}}$$^{n}=\bf{1}$$^{q}\oplus \bf{2\oplus \widetilde{3}\oplus \widetilde{3}^{\prime}}$\\
$\bf{\widetilde{3}}$$^m\otimes \bf{\overline{3}}$$^{n}=\bf{3}$$^{p}\oplus \bf{6}$\\
$\bf{\overline{3}}$$^m\otimes \bf{\overline{3}}$$^{n}=\bf{3\oplus 3^{\prime}\oplus \widetilde{3}}$$^{p}$\\
$\bf{3}$$^{m}\otimes \bf{6=3\oplus \widetilde{3}\oplus 3^{\prime}\oplus \widetilde{3}^{\prime}\oplus 6}$\\
$\bf{\widetilde{3}}$$^{m}\otimes \bf{6=3\oplus \overline{3}\oplus 3^{\prime}\oplus \overline{3}^{\prime}\oplus 6}$\\
$\bf{\overline{3}}$$^{m}\bf{\otimes 6=\widetilde{3}\oplus \overline{3}\oplus \widetilde{3}^{\prime}\oplus \overline{3}^{\prime}\oplus 6}$\\
$\bf{6\otimes 6=1\oplus 1^{\prime}\oplus 2\oplus 2 \oplus 3\oplus 3^{\prime}\oplus \widetilde{3}\oplus \widetilde{3}^{\prime}\oplus \overline{3}\oplus \overline{3}^{\prime}\oplus 6\oplus 6}$\\
\hline
\end{tabular}
\caption{The Kronecker Products of $\Delta(96)$ where $m,n=0,1$ count the number of primes on their corresponding representation, $p$ is equal to ``$\prime$'' if $m+n$ is even and nothing if $m+n$ is odd, and $q$ is equal to ``$\prime$'' if $m+n$ is odd and nothing if $m+n$ is even.}
\label{table:kronecker}
\end{table}

\subsection{Presentations of ${\boldsymbol{\Delta(96)}}$}

As discussed in the previous section, $\Delta(96)\cong (Z_4\times Z_4)\rtimes
S_3$.  As a result of this, $\Delta(96)$ can be generated by four
generators $a$, $b$, $c$ and $d$ subject to the rules \cite{Escobar:2008vc}:  
\be
\label{subgroup}
a^3=b^2=(ab)^2=c^4=d^4=1\ ,~cd=dc,
\ee
\be
\label{semidirect}
aca^{-1}=c^{-1}d^{-1},   \qquad      ada^{-1}=c\ ,~\,~~bcb^{-1}=d^{-1},  \qquad  ~\:~~~   bdb^{-1}=c^{-1} \ .
\ee
The generators (\textit{i.e.} $a$, $b$, $c$, and $d$) along with the rules
given above define a \textit{presentation} of $\Delta(96)$.  Notice that the
generators $a$ and $b$ define the $S_3$ subgroup of $\Delta(96)$ whereas the
generators $c$ and $d$ define the Abelian $Z_4\times Z_4$ (normal) subgroup of
$\Delta(96)$ [see Eq.~(\ref{subgroup})].  The other
relations [Eq.~(\ref{semidirect})] are consequences
of the semidirect product.  Furthermore, it is possible to relate $a$, $b$,
$c$, and $d$ to a smaller set of generators for $\Delta(96)$.  Define these
``new'' generators as $X$ and $Y$. The identities relating these two sets of generators (and thus presentations) for
$\Delta(96)$ are given in Ref.~\cite{Delta(96)orig1} as 
\be
a=Y^5XY^4\ ,\qquad
b=XY^2XY^5\, \qquad
c=XY^2XY^4\, \qquad
d=XY^2XY^6 \ .
\ee
Multiplying various combinations of $a$, $b$, $c$ and $d$ and their inverses
yields the relations for $X$ and $Y$ in terms of $a$, $b$, $c$, and $d$: 
\begin{eqnarray}
Y&=&c^{-1}b~=~bd \ , \\\nonumber
Y^2&=&c^{-1}d\ ,\\\nonumber
XY^5&=&ca^{-1}\ ,\\\nonumber
X&=&(ca^{-1})(c^{-1}b)(c^{-1}d)~=~d^{-1}ca^{-1}bc^{-1}d\ .
\end{eqnarray}
Notice that $X$ was derived by multiplying $XY^5$ on the right by $YY^2=Y^3$
and simplified using the relations in
Eqs.~(\ref{subgroup})-(\ref{semidirect}).  Then, using the preceding
definitions and the arithmetic in Eqs.~(\ref{subgroup})-(\ref{semidirect}),
it is straightforward to show that $XY=ddca^{-1}dd$ and
$XY^{-1}XY=ddac^{-1}dd$.  Then, further calculation reveals the presentation
for $\Delta(96)$ given in terms of the generators $X$ and $Y$ put forth in
Ref.~\cite{Delta(96)orig1}, 
\begin{equation}
X^2=Y^8=(XY)^3=(XY^{-1}XY)^3=1 \ .
\end{equation}
Now, we wish the generator $XY$ to be diagonal.  Ergo, there exists one more
transformation to preform.  Letting $T=XY$ and $U=X$ implies that $UT=XXY=Y$
and $XY^{-1}XY=U(UT)^{-1}UUT=UT^{-1}U^{-1}UUT=UT^{-1}UT$.  Therefore, the
presentation that will be used throughout the rest of this work is 
\begin{equation}
U^2=T^3=(UT)^8=(UT^{-1}UT)^3=1\ .\label{presenta}
\end{equation}
With this new, simpler presentation, the next task is to explicitly calculate
the generators for each irreducible representation of $\Delta(96)$.

\subsection{Generator Representations}
In the previous section, it was shown that a faithful representation of
$\Delta(96)$ can be generated by two generators, $U$ and $T$, satisfying the
presentation rules of Eq.~\eqref{presenta}.
With this in mind, we turn to Ref.~\cite{DingDelta(96)} which lists the
generators of $\Delta(96)$ derived from Ref.~\cite{Escobar:2008vc}, in a
particularly useful basis for flavour model building.  However, instead of
listing the generators as $a$, $b$, $c$, and $d=bc^3b$, as both
Refs.~\cite{DingDelta(96),Escobar:2008vc} do, we list them in the ``canonical''
$S$, $T$, and $U$ basis of $S_4$ (and $A_4$), see
e.g. \cite{KingLuhnS4}.\footnote{Notice here that $S=U(UT)^4U(UT)^4$ is not
required to generate $\Delta(96)$.  Yet, we list it here to draw analogy
with previous works using discrete flavour symmetries.}   
It turns out that the analogous $S_4$ generators, as they are written, are not useful for $\Delta(96)$
model building, as they require a permutation of $e$ and $\mu$ in the
flavour triplet to obtain a phenomenologically viable prediction for the
reactor angle.  To resolve this issue, an extra 1-2 permutation is applied to
the generators.  This permutation only affects the $U$ and $T$ generators.  As
a result of this extra transformation, one obtains a more ``natural'' basis
for $\Delta(96)$ model building in which the flavour triplet is the
aesthetically pleasing $(e, \mu, \tau)$.  However, this permutation comes at a
cost, the ${\bf 6}$ dimensional irreducible representation must be changed as
well from what is given in Ref.~\cite{DingDelta(96)}.  Furthermore, we have
also changed the basis of the doublet irreducible representation
given in Ref.~\cite{DingDelta(96)} to eliminate unnecessary factors of
$\omega$ in the CG coefficients involving the ${\bf 2}$.
Applying the preceding discussion to the triplet ${\bf 3}$, we see that 
\begin{eqnarray}\nonumber
s_3= \frac{1}{3} \left(
\begin{array}{ccc}
 -1 & 2 & 2 \\
 2 & -1 & 2 \\
 2 & 2 & -1
\end{array}\right), \;~~
u_3=\frac{1}{3} \left(
\begin{array}{ccc}
 -1+\sqrt{3} & -1-\sqrt{3} & -1 \\
 -1-\sqrt{3} & -1 & -1+\sqrt{3} \\
 -1 & -1+\sqrt{3} & -1-\sqrt{3}
\end{array}
\right),\\
s_3u_3=\frac{1}{3}\left(
\begin{array}{ccc}
 -1-\sqrt{3} & \sqrt{3}-1 & -1 \\
 \sqrt{3}-1 & -1 & -1-\sqrt{3} \\
 -1 & -1-\sqrt{3} & \sqrt{3}-1
\end{array}
\right),~~\text{ and }~~
t_3=\left(
\begin{array}{ccc}
 \omega ^2 & 0 & 0 \\
 0 & 1 & 0 \\
 0 & 0 & \omega 
\end{array}
\right).\, 
\end{eqnarray}
With these definitions for the generators of the irreducible
representation ${\bf 3}$, we proceed by listing the generators for the other irreducible
representations of $\Delta(96)$:

\vspace{-2mm}

\begin{eqnarray}
\label{irreps}
\begin{array}{cccc}
   & S & T & U \\[2mm]
\bf{1}:  & 1 & 1 & 1 \\[1mm]
\bf{1^{\prime}}:& 1 &1 & -1 \\
\bf{2}: &  \mathcal{I}_{2\times2} &\left(
\begin{array}{cc}
 \omega  & 0 \\
 0 & \omega^2 
\end{array}
\right)&\left(
\begin{array}{cc}
 0 & 1 \\
 1 & 0
\end{array}\right)\\
\bf{3}: & s_3& t_3 &u_3\\[1mm]
\bf{\overline{3}}:&s_3&t_3^*&u_3\\[1mm]
\bf{3^{\prime}}:& s_3& t_3&-u_3\\[1mm]
\bf{\overline{3}^{\prime}}:&s _3&t_3^*&-u_3\\[1mm]
\bf{\widetilde{3}}:& \mathcal{I}_{3\times3}&t_3&v s_3\\[1mm]
\bf{\widetilde{3}^{\prime}}:& \mathcal{I}_{3\times3}&t_3&-v s_3\\[1mm]
\bf{6}: &\left(
\begin{array}{cc}
 s_3&0\\
 0& s_3
\end{array}
\right)&
\left(
\begin{array}{cc}
t_3&0\\
0&t_3
\end{array}\right)&
\left(
\begin{array}{cc}
0&w\\
w^*&0
\end{array}\right)
\end{array}
\end{eqnarray}
where 
\begin{eqnarray}
v=-\left(
\begin{array}{ccc}
 0 & 0 & 1 \\
 0 & 1 & 0 \\
 1 & 0 & 0
\end{array}
\right)~~ \text{and } ~~
w=\frac{1}{3} \left(
\begin{array}{ccc}
 1+i & 1+i & 1-2 i \\
 1+i & 1-2 i & 1+i \\
 1-2 i & 1+i & 1+i
\end{array}
\right).
\end{eqnarray}
Sometimes we shall refer to the triplet ${\bf 3}$ generators simply as
$S=s_3$, $T=t_3$, $U=u_3$ (as is done in the Introduction, for example).
Notice that in Eq.~(\ref{irreps}) the ``$S$'' generator is an identity matrix
for the representations $\bf{1}$, $\bf{1^{\prime}}$, ${\bf 2}$, ${\bf \widetilde{3}}$, and
${\bf \widetilde{3}^{\prime}}$.  This is due to the aforementioned fact that
these are unfaithful representations of $\Delta(96)$.  These representations
are unable to generate the full $\Delta(96)$ symmetry.  In fact the
representations  $\bf{1}$, $\bf{1^{\prime}}$, ${\bf 2}$, ${\bf \widetilde{3}}$, and 
${\bf \widetilde{3}^{\prime}}$  generate groups isomorphic to
the \textit{trivial} group, $Z_2$, $S_3\cong\Delta(6)$, $S_4\cong\Delta(24)$, and $S_4\cong\Delta(24)$,
respectively.  Yet, the urge to claim these unfaithful representations as
irrelevant must be put aside when looking to calculate a complete list of CG
coefficients of $\Delta(96)$, which is the next step towards understanding
this group.

\subsection{$\boldsymbol{\Delta(96)}$ Clebsch-Gordan Coefficients}

In this section, we list the CG coefficients derived from the basis given in
the previous section. All CG coefficients are reported in the form $a\otimes
b$, where the $a_i$ are from the representation on the left of the product,
and the $b_j$ are from the representation on the right of the product.  Notice
that from ${\bf 3\otimes 3}$ onward, a single set of CG coefficients yields
the results for two sets of tensor product decompositions.  The tensor product
on the left has its representations labelled on the left, and the tensor
product on the right has its representations labelled on the right.  In
addition to these guidelines, note that the subscripts ``${\bf s}$'' and ``${\bf a}$'' denote symmetric and anti-symmetric, respectively. 

\newpage~\newline\newline\newline\newline

\begin{eqnarray}\nonumber
\begin{array}{|cc|cc|}\hline
\boxed{\bf{1^{\prime} \otimes 2=2}}& & &\boxed{\bf{1^{\prime} \otimes 6=6}}\\
&&&\\
\bf{2}\sim\left(
\begin{array}{c}
 a_1 b_1 \\
 -a_1 b_2
\end{array}
\right)&&&
\bf{6}\sim \left(
\begin{array}{c}
 a_1 b_1 \\
 a_1 b_2 \\
 a_1 b_3 \\
 -a_1 b_4 \\
 -a_1 b_5 \\
 -a_1 b_6
\end{array}
\right)\\
&&&\\\hline
%\end{array}
%\end{eqnarray}
%\begin{eqnarray}\nonumber
%\begin{array}{cccc}
\boxed{\bf{1^{\prime}} \otimes r=r^{\prime}~\text{for}~\bf{r=3},~\bf{\widetilde{3}},~\text{or}~ \bf{\overline{3}}}&&&\boxed{\bf{1^{\prime} \otimes r^{\prime}=r} ~\text{for}~\bf{r=3},~\bf{\widetilde{3}},~\text{or}~ \bf{\overline{3}}}\\
&&&\\
\bf{r^{\prime}}\sim\left(
\begin{array}{c}
 a_1 b_1 \\
 a_1 b_2 \\
 a_1 b_3
\end{array}
\right)
&&&
\bf{r}\sim\left(
\begin{array}{c}
 a_1 b_1 \\
 a_1 b_2 \\
 a_1 b_3
\end{array}
\right)\\
&&&\\\hline
%\end{array}
%\end{eqnarray}
%\begin{eqnarray}\nonumber
%\begin{array}{cccc}
\boxed{\bf{2 \otimes 2=1_s\oplus 1_a^{\prime}\oplus 2_s}}&&&\boxed{\bf{2\otimes 6=6_1\oplus 6_2}}\\
&&&\\
\bf{1_s}\sim  \text{$a_1 b_2+a_2 b_1$}&&&
\bf{6_1}\sim\left(
\begin{array}{c}
 a_1 b_3 \\
 a_1 b_1 \\
 a_1 b_2 \\
 a_2 b_5 \\
 a_2 b_6 \\
 a_2 b_4
\end{array}
\right)\\
&&&\\
\bf{1^{\prime}_a}\sim\text{$ a_1 b_2-a_2 b_1$}&&&\\
\bf{2_s}\sim \left(
\begin{array}{c}
 a_2 b_2 \\
 a_1 b_1
\end{array}
\right)&&&
\bf{6_2}\sim
\left(
\begin{array}{c}
 a_2 b_2 \\
 a_2 b_3 \\
 a_2 b_1 \\
 a_1 b_6 \\
 a_1 b_4 \\
 a_1 b_5
\end{array}
\right)\\
&&&\\\hline
\end{array}
\end{eqnarray}

\newpage
~\newline
\newline
\newline
\newline
\newline

\begin{eqnarray}\nonumber
\begin{array}{|cc|cc|}\hline
\boxed{\bf{2\otimes r}=\bf{r \oplus r^{\prime}}~\text{when}~\bf{r=3}~\text{or}~\bf{\widetilde{3}}}&&&\boxed{\bf{2\otimes r^{\prime}}=\bf{r \oplus r^{\prime}}~\text{when}~\bf{r=3}~\text{or}~\bf{\widetilde{3}}}\\
&&&\\
\bf{r}\sim\left(
\begin{array}{c}
 a_2 b_2+a_1 b_3 \\
 a_2 b_3+a_1 b_1 \\
 a_2 b_1+a_1 b_2
\end{array}
\right)&&&
\bf{r}\sim\left(
\begin{array}{c}
 a_2 b_2-a_1 b_3 \\
 a_2 b_3-a_1 b_1 \\
 a_2 b_1-a_1 b_2
\end{array}
\right)\\
&&&\\
\bf{r}^{\prime}\sim\left(
\begin{array}{c}
 a_2 b_2-a_1 b_3 \\
 a_2 b_3-a_1 b_1 \\
 a_2 b_1-a_1 b_2
\end{array}
\right)&&&
\bf{r}^{\prime}\sim\left(
\begin{array}{c}
 a_2 b_2+a_1 b_3 \\
 a_2 b_3+a_1 b_1 \\
 a_2 b_1+a_1 b_2
\end{array}
\right)\\&&&\\\hline
%\end{array}
%\end{eqnarray}
%\begin{eqnarray}\nonumber
%\begin{array}{cccc}
\boxed{\bf{2\otimes \overline{3}}=\overline{3}\oplus \overline{3}^{\prime}}&&&\boxed{\bf{2\otimes \overline{3}}^{\prime}=\overline{3}\oplus \overline{3}^{\prime}}\\
&&&\\
\bf{\overline{3}}\sim\left(
\begin{array}{c}
 a_1 b_2+a_2 b_3 \\
 a_1 b_3+a_2 b_1 \\
 a_1 b_1+a_2 b_2
\end{array}
\right)&&&\bf{\overline{3}}\sim\left(
\begin{array}{c}
 a_1 b_2-a_2 b_3 \\
 a_1 b_3-a_2 b_1 \\
 a_1 b_1-a_2 b_2
\end{array}
\right)\\&&&\\
\bf{\overline{3}}^{\prime}\sim\left(
\begin{array}{c}
 a_1 b_2-a_2 b_3 \\
 a_1 b_3-a_2 b_1 \\
 a_1 b_1-a_2 b_2
\end{array}
\right)&&&\bf{\overline{3}}^{\prime}\sim\left(
\begin{array}{c}
 a_1 b_2+a_2 b_3 \\
 a_1 b_3+a_2 b_1 \\
 a_1 b_1+a_2 b_2
\end{array}
\right)\\
&&&\\\hline
%\end{array}
%\end{eqnarray}
%\begin{eqnarray}\nonumber
%\begin{array}{cccc}
\boxed{\bf{3\otimes 3=3^{\prime} \otimes 3^{\prime}=\widetilde{3}_s^{\prime}\oplus \overline{3}_a\oplus \overline{3}_s^{\prime}}}&&&\boxed{\bf{3^{\prime} \otimes 3 =\widetilde{3}\oplus \overline{3}\oplus \overline{3}^{\prime}}}\\
&&&\\
\bf{\widetilde{3}^{\prime}_s}\sim\left(
\begin{array}{c}
 a_1 b_2+a_2 b_1+a_3 b_3 \\
 a_1 b_3+a_2 b_2+a_3 b_1 \\
 a_1 b_1+a_2 b_3+a_3 b_2
\end{array}\right)&&&\bf{\widetilde{3}}\sim\left(
\begin{array}{c}
 a_1 b_2+a_2 b_1+a_3 b_3 \\
 a_1 b_3+a_2 b_2+a_3 b_1 \\
 a_1 b_1+a_2 b_3+a_3 b_2
\end{array}
\right)\\
&&&\\
\bf{\overline{3}_a}\sim\left(
\begin{array}{c}
 a_2 b_3-a_3 b_2 \\
 a_3 b_1-a_1 b_3 \\
 a_1 b_2-a_2 b_1
\end{array}
\right)&&&\bf{\overline{3}}\sim \left(
\begin{array}{c}
 -2 a_1 b_1+a_2 b_3+a_3 b_2 \\
 a_1 b_3-2 a_2 b_2+ a_3 b_1\\
 a_1 b_2+a_2 b_1-2 a_3 b_3
\end{array}
\right)\\
&&&\\
\bf{\overline{3}^{\prime}_s}\sim\left(
\begin{array}{c}
 -2 a_1 b_1+a_2 b_3+a_3 b_2 \\
 a_1 b_3-2 a_2 b_2+ a_3 b_1\\
 a_1 b_2+a_2 b_1-2 a_3 b_3
\end{array}
\right)&&&\bf{\overline{3}^{\prime}}\sim\left(
\begin{array}{c}
 a_2 b_3-a_3 b_2 \\
 a_3 b_1-a_1 b_3 \\
 a_1 b_2-a_2 b_1
\end{array}
\right)\\
&&&\\\hline
\end{array}
\end{eqnarray}
\newpage

\begin{eqnarray}\nonumber
\begin{array}{|cc|cc|}\hline
\boxed{\bf{3 \otimes \widetilde{3}=3^{\prime} \otimes \widetilde{3}^{\prime}=\overline{3}^{\prime}\oplus 6}}&&& \boxed{\bf{3 \otimes \widetilde{3}^{\prime}=3^{\prime} \otimes \widetilde{3}=\overline{3}\oplus 6}}\\
&&&\\
\bf{\overline{3}^{\prime}}\sim\left(
\begin{array}{c}
 a_1 b_1+a_2 b_3+a_3 b_2 \\
 a_1 b_3+a_2 b_2+a_3 b_1 \\
 a_1 b_2+a_2 b_1+a_3 b_3
\end{array}
\right)&&& \bf{\overline{3}}\sim\left(
\begin{array}{c}
 a_1 b_1+a_2 b_3+a_3 b_2 \\
 a_1 b_3+a_2 b_2+a_3 b_1 \\
 a_1 b_2+a_2 b_1+a_3 b_3
\end{array}
\right)\\
&&&\\
\bf{6}\sim\left(
\begin{array}{c}
 a_1 b_2+\omega ^2 a_2 b_1 +\omega a_3 b_3   \\
 \omega a_1 b_3  +a_2 b_2+\omega ^2 a_3 b_1  \\
 \omega ^2a_1 b_1 +\omega a_2 b_3  +a_3 b_2 \\
 -a_1 b_2-\omega a_2 b_1-\omega ^2 a_3 b_3    \\
 -\omega ^2 a_1 b_3 -a_2 b_2-\omega a_3 b_1   \\
 -\omega a_1 b_1-\omega ^2 a_2 b_3   -a_3 b_2
\end{array}
\right)&&&\bf{6}\sim\left(
\begin{array}{c}
 a_1 b_2+\omega ^2 a_2 b_1 +\omega a_3 b_3   \\
 \omega a_1 b_3  +a_2 b_2+\omega ^2 a_3 b_1  \\
 \omega ^2a_1 b_1 +\omega a_2 b_3  +a_3 b_2 \\
 a_1 b_2+\omega a_2 b_1+\omega ^2 a_3 b_3    \\
 \omega ^2 a_1 b_3 +a_2 b_2+\omega a_3 b_1   \\
 \omega a_1 b_1+\omega ^2 a_2 b_3 +a_3 b_2
\end{array}
\right)\\
&&&\\\hline
%\end{array}
%\end{eqnarray}
%\begin{eqnarray}\nonumber
%\begin{array}{cccc}
\boxed{\bf{3 \otimes \overline{3}=3^{\prime} \otimes \overline{3}^{\prime}=1\oplus 2\oplus 6}} &&& \boxed{\bf{3\otimes \overline{3}^{\prime}=3^{\prime}\otimes \overline{3}=1^{\prime}\oplus 2\oplus 6}}\\
&&&\\
\bf{1}\sim \text{$a_1 b_1+a_2 b_2+a_3 b_3$}&&&\bf{1}^{\prime}\sim \text{$a_1 b_1+a_2 b_2+a_3 b_3$}\\
&&&\\
\bf{2}\sim\left(
\begin{array}{c}
 a_1 b_3+a_2 b_1+a_3 b_2 \\
 a_1 b_2+a_2 b_3+a_3 b_1
\end{array}
\right)&&&\bf{2}\sim\left(
\begin{array}{c}
 a_1 b_3+a_2 b_1+a_3 b_2 \\
 -a_1 b_2-a_2 b_3-a_3 b_1
\end{array}
\right)\\
&&&\\
\bf{6}\sim \left(
\begin{array}{c}
 a_1 b_2+\omega ^2a_2 b_3 +\omega a_3 b_1   \\
 \omega ^2a_1 b_1 +\omega a_2 b_2  +a_3 b_3 \\
 \omega a_1 b_3+a_2 b_1  +\omega ^2a_3 b_2  \\
 -\omega ^2a_1 b_2  -a_2 b_3-\omega a_3 b_1  \\
 -a_1 b_1-\omega a_2 b_2-\omega ^2a_3 b_3    \\
 -\omega a_1 b_3-\omega ^2a_2 b_1   -a_3 b_2
\end{array}
\right)&&&\bf{6}\sim \left(
\begin{array}{c}
 a_1 b_2+\omega ^2a_2 b_3 +\omega a_3 b_1   \\
 \omega ^2a_1 b_1 +\omega a_2 b_2  +a_3 b_3 \\
 \omega a_1 b_3+a_2 b_1  +\omega ^2a_3 b_2  \\
 \omega ^2a_1 b_2  +a_2 b_3+\omega a_3 b_1  \\
 a_1 b_1+\omega a_2 b_2+\omega ^2a_3 b_3    \\
 \omega a_1 b_3+\omega ^2a_2 b_1   +a_3 b_2
\end{array}
\right)\\
&&&\\\hline
%\end{array}
%\end{eqnarray}\\
%\begin{eqnarray}\nonumber
%\begin{array}{cccc}
\boxed{\bf{\widetilde{3}\otimes \widetilde{3} =\widetilde{3}^{\prime}\otimes \widetilde{3}^{\prime}= 1_s\oplus 2_s\oplus \widetilde{3}_a\oplus \widetilde{3}_s^{\prime}}}&&& \boxed{\bf{\widetilde{3}^{\prime} \otimes \widetilde{3} = 1^{\prime}\oplus 2\oplus \widetilde{3}^{\prime}\oplus \widetilde{3}}}\\
&&&\\
\bf{1_s}\sim \text{$a_1 b_3+a_2 b_2+a_3 b_1$}&&&\bf{1^{\prime}}\sim \text{$a_1 b_3+a_2 b_2+a_3 b_1$}\\
&&&\\
\bf{2_s}\sim\left(
\begin{array}{c}
 a_1 b_1+a_2 b_3+a_3 b_2 \\
 a_1 b_2+a_2 b_1+a_3 b_3
\end{array}
\right)&&& \bf{2}\sim\left(
\begin{array}{c}
 a_1 b_1+a_2 b_3+a_3 b_2 \\
 -a_1 b_2-a_2 b_1-a_3 b_3
\end{array}
\right)\\
&&&\\
\bf{\widetilde{3}_a}\sim\left(
\begin{array}{c}
 a_1 b_2-a_2 b_1 \\
 a_3 b_1-a_1 b_3\\
 a_2 b_3-a_3 b_2 \\ 
\end{array}
\right)&&&\bf{\widetilde{3}^{\prime}}\sim\left(
\begin{array}{c}
 a_1 b_2-a_2 b_1 \\
 a_3 b_1-a_1 b_3\\
 a_2 b_3-a_3 b_2 \\ 
\end{array}
\right)\\
&&&\\
\bf{\widetilde{3}^{\prime}_s}\sim\left(
\begin{array}{c}
 a_1 b_2+a_2 b_1-2 a_3 b_3 \\
 a_1 b_3-2a_2 b_2+ a_3 b_1 \\
 -2 a_1 b_1+ a_2 b_3+a_3 b_2
\end{array}
\right)&&&\bf{\widetilde{3}}\sim\left(
\begin{array}{c}
 a_1 b_2+a_2 b_1-2 a_3 b_3 \\
 a_1 b_3-2a_2 b_2+ a_3 b_1 \\
 -2 a_1 b_1+ a_2 b_3+a_3 b_2
\end{array}
\right)\\
&&&\\\hline
\end{array}
\end{eqnarray}

\newpage
~\newline\newline\newline\newline\newline\newline

\begin{eqnarray}\nonumber
\begin{array}{|cc|cc|}\hline
\boxed{\bf{\widetilde{3}\otimes \overline{3}=\widetilde{3}^{\prime}\otimes \overline{3}^{\prime}= 3^{\prime} \oplus 6}}&&& \boxed{ \bf{\widetilde{3} \otimes \overline{3}^{\prime} = \widetilde{3}^{\prime}\otimes \overline{3}  =3 \oplus 6}}\\
&&&\\\bf{3^{\prime}}\sim\left(
\begin{array}{c}
 a_1 b_2+a_2 b_3+a_3 b_1 \\
 a_1 b_1+a_2 b_2+a_3 b_3 \\
 a_1 b_3+a_2 b_1+a_3 b_2
\end{array}
\right)&&&\bf{3}\sim\left(
\begin{array}{c}
 a_1 b_2+a_2 b_3+a_3 b_1 \\
 a_1 b_1+a_2 b_2+a_3 b_3 \\
 a_1 b_3+a_2 b_1+a_3 b_2
\end{array}
\right)\\
&&&\\
\bf{6}\sim \left(
\begin{array}{c}
 a_1 b_2+\omega ^2 a_2 b_3 +\omega a_3 b_1   \\
 a_1 b_1+\omega ^2 a_2 b_2 +\omega a_3 b_3   \\
 a_1 b_3+\omega ^2a_2 b_1 +\omega a_3 b_2   \\
 -\omega a_1 b_2-\omega ^2a_2 b_3   -a_3 b_1 \\
 -\omega a_1 b_1-\omega ^2a_2 b_2   -a_3 b_3 \\
  -\omega a_1 b_3-\omega ^2a_2 b_1  -a_3 b_2
\end{array}
\right)&&&\bf{6}\sim\left(
\begin{array}{c}
 a_1 b_2+\omega ^2 a_2 b_3 +\omega a_3 b_1   \\
 a_1 b_1+\omega ^2 a_2 b_2 +\omega a_3 b_3   \\
 a_1 b_3+\omega ^2a_2 b_1 +\omega a_3 b_2   \\
 \omega a_1 b_2+\omega ^2a_2 b_3   +a_3 b_1 \\
 \omega a_1 b_1+\omega ^2a_2 b_2   +a_3 b_3 \\
  \omega a_1 b_3+\omega ^2a_2 b_1  +a_3 b_2
\end{array}
\right)\\
&&&\\\hline
%\end{array}
%\end{eqnarray}
%\begin{eqnarray}\nonumber
%\begin{array}{cccc}
\boxed{\bf{\overline{3}\otimes\overline{3}=\overline{3}^{\prime}\otimes\overline{3}^{\prime}=    3_a \oplus 3_s^{\prime} \oplus \widetilde{3}_s^{\prime}}}&&&\boxed{\bf{\overline{3}^{\prime} \otimes \overline{3} =  3^{\prime}\oplus 3  \oplus \widetilde{3}}}\\
&&&\\\bf{3_a}\sim\left(
\begin{array}{c}
 a_2 b_3-a_3 b_2 \\
 a_3 b_1-a_1 b_3 \\
 a_1 b_2-a_2 b_1
\end{array}
\right)&&&\bf{3^{\prime}}\sim\left(
\begin{array}{c}
 a_2 b_3-a_3 b_2 \\
 a_3 b_1-a_1 b_3 \\
 a_1 b_2-a_2 b_1
\end{array}
\right)\\
&&&\\\bf{3^{\prime}_s}\sim\left(
\begin{array}{c}
 -2 a_1 b_1+a_2 b_3+a_3 b_2 \\
 a_1 b_3-2 a_2 b_2+a_3 b_1 \\
 a_1 b_2+a_2 b_1-2 a_3 b_3
\end{array}
\right)&&&\bf{3}\sim\left(
\begin{array}{c}
 -2 a_1 b_1+a_2 b_3+a_3 b_2 \\
 a_1 b_3-2 a_2 b_2+a_3 b_1 \\
 a_1 b_2+a_2 b_1-2 a_3 b_3
\end{array}
\right)\\
&&&\\\bf{\widetilde{3}^{\prime}_s}\sim\left(
\begin{array}{c}
 a_1 b_1+a_2 b_3+a_3 b_2 \\
 a_1 b_3+a_2 b_2+a_3 b_1 \\
 a_1 b_2+a_2 b_1+a_3 b_3
\end{array}
\right)&&&\bf{\widetilde{3}}\sim\left(
\begin{array}{c}
 a_1 b_1+a_2 b_3+a_3 b_2 \\
 a_1 b_3+a_2 b_2+a_3 b_1 \\
 a_1 b_2+a_2 b_1+a_3 b_3
\end{array}
\right)\\
&&&\\\hline
\end{array}
\end{eqnarray}

\newpage
~\newline
\newline
\newline
\newline
\newline
\newline
\newline
\newline
\newline
\begin{centering}
\fbox{
\begin{minipage}[h]{5.5 in}
%\begin{eqnarray}\nonumber
%\begin{array}{cc|cc}
%\boxed{\bf{3\otimes 6 = 3 \oplus \widetilde{3} \oplus 3^{\prime} \oplus \widetilde{3}^{\prime} \oplus 6}}&&&\boxed{\bf{3^{\prime} \otimes 6 = 3 \oplus \widetilde{3} \oplus 3^{\prime} \oplus \widetilde{3}^{\prime} \oplus 6}}
%\end{array}
%\end{eqnarray}
\fbox{$\bf{3\otimes 6 = 3 \oplus \widetilde{3} \oplus 3^{\prime} \oplus \widetilde{3}^{\prime} \oplus 6}$}\hspace{.58in}$\Big |$\hspace{.6in} \fbox{$\bf{3^{\prime} \otimes 6 =  3^{\prime} \oplus \widetilde{3}^{\prime} \oplus 3 \oplus \widetilde{3}\oplus  6}$}
\begin{eqnarray}\nonumber
\begin{array}{c}
\bf{3}\longrightarrow\left(
\begin{array}{c}
 a_1 b_2-\omega ^2a_1 b_5 +\omega ^2a_2 b_1-a_2 b_4   +\omega a_3 b_3  -\omega a_3 b_6  \\
 \omega ^2a_1 b_3-a_1 b_6 +\omega a_2 b_2  -\omega a_2 b_5+a_3 b_1-\omega ^2a_3 b_4    \\
 \omega a_1 b_1  -\omega a_1 b_4+a_2 b_3-\omega ^2a_2 b_6+\omega ^2a_3 b_2  -a_3 b_5
\end{array}
\right)\longleftarrow\bf{ 3^{\prime}}\\
\\
%\end{eqnarray}
%\begin{eqnarray}
\bf{\widetilde{3}}\longrightarrow \left(
\begin{array}{c}
 a_1 b_2-\omega ^2 a_1 b_5+a_2 b_1-\omega ^2 a_2 b_4  +a_3 b_3-\omega ^2a_3 b_6  \\
 \omega a_1 b_3-\omega a_1 b_6 +\omega a_2 b_2-\omega a_2 b_5+\omega a_3 b_1 -\omega a_3 b_4     \\
 \omega ^2a_1 b_1-a_1 b_4 +\omega ^2 a_3 b_2 -a_3 b_5+\omega ^2 a_2 b_3 -a_2 b_6
\end{array}
\right)\longleftarrow \bf{\widetilde{3}^{\prime}}\\\\
%\end{array}
%\end{eqnarray}
%\begin{eqnarray}
\bf{3^{\prime}}\longrightarrow\left(
\begin{array}{c}
 a_1 b_2+\omega ^2a_1 b_5 +\omega ^2a_2 b_1+a_2 b_4   +\omega a_3 b_3  +\omega a_3 b_6  \\
 \omega ^2a_1 b_3+a_1 b_6+\omega a_2 b_2  +\omega a_2 b_5 +a_3 b_1+\omega ^2a_3 b_4    \\
 \omega a_1 b_1  +\omega a_1 b_4+a_2 b_3+\omega ^2a_2 b_6+\omega ^2a_3 b_2  +a_3 b_5
\end{array}
\right)\longleftarrow \bf{3}\\\\
%\end{eqnarray}
%\begin{eqnarray}
\bf{\widetilde{3}^{\prime}}\longrightarrow\left(
\begin{array}{c}
 a_1 b_2+\omega ^2 a_1 b_5+a_2 b_1+\omega ^2 a_2 b_4  +a_3 b_3+\omega ^2a_3 b_6  \\
 \omega a_1 b_3+\omega a_1 b_6+\omega a_2 b_2+\omega a_2 b_5 +\omega a_3 b_1 +\omega a_3 b_4     \\
 \omega ^2a_1 b_1+a_1 b_4 +\omega ^2 a_2 b_3 +a_2 b_6+\omega ^2 a_3 b_2 +a_3 b_5
\end{array}
\right)\longleftarrow\bf{\widetilde{3}}
\end{array}
\end{eqnarray}
\begin{eqnarray}\nonumber
\begin{array}{c|c}
\bf{6}\sim\left(
\begin{array}{c}
 a_1 b_2 +\omega a_2 b_1+\omega ^2a_3 b_3   \\
 \omega a_1 b_3 +\omega ^2a_2 b_2  +a_3 b_1 \\
 \omega ^2a_1 b_1   +a_2 b_3 +\omega a_3 b_2\\
 -\omega a_1 b_5  -a_2 b_4 -\omega ^2a_3 b_6 \\
 -a_1 b_6 -\omega ^2a_2 b_5 -\omega a_3 b_4  \\
 -\omega ^2a_1 b_4 -\omega a_2 b_6  -a_3 b_5
\end{array}
\right)&\bf{6}\sim\left(
\begin{array}{c}
 a_1 b_2 +\omega a_2 b_1+\omega ^2a_3 b_3   \\
 \omega a_1 b_3 +\omega ^2a_2 b_2  +a_3 b_1 \\
 \omega ^2a_1 b_1   +a_2 b_3 +\omega a_3 b_2\\
 \omega a_1 b_5  +a_2 b_4 +\omega ^2a_3 b_6 \\
 a_1 b_6 +\omega ^2a_2 b_5 +\omega a_3 b_4  \\
 \omega ^2a_1 b_4 +\omega a_2 b_6  +a_3 b_5
\end{array}
\right)\\
&\\
\end{array}
\end{eqnarray}
\end{minipage}}
\end{centering}
\newpage

~\newline
\newline
\newline
\newline
\newline
\newline
\newline
\newline
\newline
\begin{centering}
\fbox{
\begin{minipage}[h]{5.5 in}
\fbox{$\bf{\widetilde{3} \otimes 6 = 3 \oplus \overline{3} \oplus 3^{\prime} \oplus \overline{3}^{\prime} \oplus 6}$}\hspace{.58in}$\Big|$\hspace{.6in}\fbox{ $\bf{\widetilde{3}^{\prime} \otimes 6 =3^{\prime} \oplus \overline{3}^{\prime} \oplus 3 \oplus \overline{3} \oplus  6}$}
\begin{eqnarray}\nonumber
\begin{array}{c}
\bf{3}\longrightarrow\left(
\begin{array}{c}
 a_1 b_2-\omega ^2a_1 b_5+\omega a_2 b_1 -\omega a_2 b_4 +\omega ^2a_3 b_3   -a_3 b_6 \\
 a_1 b_3-\omega ^2a_1 b_6+\omega a_2 b_2  -\omega a_2 b_5 +\omega ^2a_3 b_1   -a_3 b_4 \\
 a_1 b_1-\omega ^2a_1 b_4+\omega a_2 b_3 -\omega a_2 b_6+\omega ^2a_3 b_2     -a_3 b_5
\end{array}
\right)\longleftarrow \bf{3^{\prime}}\\\\
%\end{eqnarray}
%\begin{eqnarray}
\bf{\overline{3}}\longrightarrow\left(
\begin{array}{c}
 a_1 b_1-\omega a_1 b_4+\omega ^2a_2 b_3 -\omega ^2a_2 b_6 +\omega a_3 b_2 -a_3 b_5 \\
 a_1 b_3-\omega a_1 b_6+\omega ^2a_2 b_2 -\omega ^2a_2 b_5 +\omega a_3 b_1  -a_3 b_4 \\
 a_1 b_2-\omega a_1 b_5+\omega ^2a_2 b_1 -\omega ^2a_2 b_4 +\omega a_3 b_3  -a_3 b_6
\end{array}
\right)\longleftarrow\bf{\overline{3}^{\prime}}\\\\
%\end{eqnarray}
%\begin{eqnarray}
\bf{3^{\prime}}\longrightarrow \left(
\begin{array}{c}
 a_1 b_2+\omega ^2a_1 b_5+\omega a_2 b_1 +\omega a_2 b_4 +\omega ^2a_3 b_3   +a_3 b_6 \\
 a_1 b_3+\omega ^2a_1 b_6+\omega a_2 b_2  +\omega a_2 b_5 +\omega ^2a_3 b_1   +a_3 b_4 \\
 a_1 b_1+\omega ^2a_1 b_4+\omega a_2 b_3 +\omega a_2 b_6+\omega ^2a_3 b_2     +a_3 b_5
\end{array}
\right)\longleftarrow \bf{3}\\\\
%\end{eqnarray}
%\begin{eqnarray}
\bf{\overline{3}^{\prime}}\longrightarrow
\left(
\begin{array}{c}
 a_1 b_1+\omega a_1 b_4+\omega ^2a_2 b_3 +\omega ^2a_2 b_6 +\omega a_3 b_2 +a_3 b_5 \\
 a_1 b_3+\omega a_1 b_6+\omega ^2a_2 b_2 +\omega ^2a_2 b_5 +\omega a_3 b_1  +a_3 b_4 \\
 a_1 b_2+\omega a_1 b_5+\omega ^2a_2 b_1 +\omega ^2a_2 b_4 +\omega a_3 b_3  +a_3 b_6
\end{array}
\right)\longleftarrow\bf{\overline{3}}
\end{array}
\end{eqnarray}
\begin{eqnarray}\nonumber
\begin{array}{c|c}
\bf{6}\sim\left(
\begin{array}{c}
 a_1 b_5+a_2 b_4+a_3 b_6 \\
 a_1 b_6+a_2 b_5 +a_3 b_4\\
 a_1 b_4+a_2 b_6 +a_3 b_5\\
 -a_1 b_2-a_2 b_1-a_3 b_3 \\
 -a_1 b_3 -a_2 b_2-a_3 b_1\\
 -a_1 b_1-a_2 b_3-a_3 b_2
\end{array}
\right)&\bf{6}\sim \left(
\begin{array}{c}
 a_1 b_5+a_2 b_4+a_3 b_6 \\
 a_1 b_6+a_2 b_5 +a_3 b_4\\
 a_1 b_4+a_2 b_6 +a_3 b_5\\
 a_1 b_2+a_2 b_1+a_3 b_3 \\
 a_1 b_3 +a_2 b_2+a_3 b_1\\
 a_1 b_1+a_2 b_3+a_3 b_2
\end{array}
\right)
\\
&\\
\end{array}
\end{eqnarray}
\end{minipage}}
\end{centering}

\newpage
~\newline
\newline
\newline
\newline
\newline
\newline
\newline
\newline
\newline
\begin{centering}
\fbox{
\begin{minipage}[h]{5.5 in}
\fbox{$ \bf{\overline{3} \otimes 6 = \widetilde{3} \oplus \overline{3} \oplus \widetilde{3}^{\prime} \oplus \overline{3}^{\prime} \oplus 6}$}\hspace{.58in}$\Big |$\hspace{.6in} \fbox{$\bf{\overline{3}^{\prime} \otimes 6 = \widetilde{3}^{\prime} \oplus \overline{3}^{\prime} \oplus\widetilde{3} \oplus \overline{3} \oplus  6}$}
\begin{eqnarray}\nonumber
\begin{array}{c}
\bf{\widetilde{3}}\longrightarrow \left(
\begin{array}{c}
 a_1 b_3-\omega a_1 b_6+a_2 b_1-\omega a_2 b_4  +a_3 b_2 -\omega a_3 b_5    \\
  \omega ^2 a_1 b_1-\omega ^2a_1 b_4 +\omega ^2a_2 b_2-\omega ^2a_2 b_5 +\omega ^2a_3 b_3 -\omega ^2a_3 b_6 \\
 \omega a_1 b_2-a_1 b_5+\omega a_2 b_3-a_2 b_6+\omega a_3 b_1-a_3 b_4
\end{array}
\right)\longleftarrow\bf{\widetilde{3}^{\prime}}\\\\
%\end{eqnarray}
%\begin{eqnarray}
\bf{\overline{3}}\longrightarrow\left(
\begin{array}{c}
 a_1 b_2-\omega ^2a_1 b_5+\omega ^2a_2 b_3-a_2 b_6   +\omega a_3 b_1  -\omega a_3 b_4  \\
 \omega ^2a_1 b_1-a_1 b_4 +\omega a_2 b_2 -\omega a_2 b_5+a_3 b_3-\omega ^2a_3 b_6     \\
 \omega a_1 b_3  -\omega a_1 b_6+a_2 b_1-\omega ^2a_2 b_4+\omega ^2a_3 b_2 -a_3 b_5
\end{array}
\right)\longleftarrow\bf{\overline{3}^{\prime}}\\\\
%\end{eqnarray}
%\begin{eqnarray}
\bf{\widetilde{3}^{\prime}}\longrightarrow\left(
\begin{array}{c}
 a_1 b_3+\omega a_1 b_6+a_2 b_1+\omega a_2 b_4  +a_3 b_2 +\omega a_3 b_5    \\
  \omega ^2 a_1 b_1+\omega ^2a_1 b_4 +\omega ^2a_2 b_2+\omega ^2a_2 b_5 +\omega ^2a_3 b_3 +\omega ^2a_3 b_6 \\
 \omega a_1 b_2+a_1 b_5+\omega a_2 b_3+a_2 b_6+\omega a_3 b_1+a_3 b_4
\end{array}
\right)\longleftarrow\bf{\widetilde{3}}\\\\
%\end{eqnarray}
%\begin{eqnarray}
\bf{\overline{3}^{\prime}}\longrightarrow\left(
\begin{array}{c}
 a_1 b_2+\omega ^2a_1 b_5+\omega ^2a_2 b_3+a_2 b_6   +\omega a_3 b_1  +\omega a_3 b_4  \\
 \omega ^2a_1 b_1+a_1 b_4 +\omega a_2 b_2 +\omega a_2 b_5+a_3 b_3+\omega ^2a_3 b_6     \\
 \omega a_1 b_3  +\omega a_1 b_6+a_2 b_1+\omega ^2a_2 b_4+\omega ^2a_3 b_2 +a_3 b_5
\end{array}
\right)\longleftarrow\bf{\overline{3}}
\end{array}
\end{eqnarray}

\begin{eqnarray}\nonumber
\begin{array}{c|c}
\bf{6}\sim\left(
\begin{array}{c}
 a_1 b_3+\omega a_2 b_1+\omega ^2a_3 b_2    \\
 \omega ^2a_1 b_1 +a_2 b_2+\omega a_3 b_3   \\
 \omega a_1 b_2+\omega ^2a_2 b_3   +a_3 b_1 \\
 -a_1 b_6-\omega ^2a_2 b_4 -\omega a_3 b_5   \\
 -\omega a_1 b_4  -a_2 b_5-\omega ^2a_3 b_6  \\
 -\omega ^2a_1 b_5 -\omega a_2 b_6  -a_3 b_4
\end{array}
\right)&\bf{6}\sim\left(
\begin{array}{c}
 a_1 b_3+\omega a_2 b_1+\omega ^2a_3 b_2    \\
 \omega ^2a_1 b_1 +a_2 b_2+\omega a_3 b_3   \\
 \omega a_1 b_2+\omega ^2a_2 b_3   +a_3 b_1 \\
 a_1 b_6+\omega ^2a_2 b_4 +\omega a_3 b_5   \\
 \omega a_1 b_4 +a_2 b_5+\omega ^2a_3 b_6  \\
 \omega ^2a_1 b_5 +\omega a_2 b_6  +a_3 b_4
\end{array}
\right)\\
&\\
\end{array}
\end{eqnarray}
\end{minipage}}
\end{centering}

\begin{centering}
\fbox{
\begin{minipage}[h]{5.5 in}
\hspace{.25in}\fbox{$\bf{6 \otimes 6 = 1_s \oplus 1_a^{\prime} \oplus 2_s \oplus 2_a \oplus 3_a\oplus \widetilde{3}_s\oplus\overline{3}_a\oplus 3^{\prime}_s\oplus \widetilde{3}_s^{\prime}\oplus \overline{3}^{\prime}_s\oplus  6_s \oplus 6_a}$}
\begin{eqnarray}\nonumber
\begin{array}{c}
%\begin{eqnarray}
\bf{1_s}\sim \text{$a_1 b_6+a_2 b_5+a_3 b_4+a_4 b_3+a_5 b_2+a_6 b_1$}\\\\
%\end{eqnarray}
%\begin{eqnarray}
\bf{1^{\prime}_a}\sim \text{$a_1 b_6+a_2 b_5+a_3 b_4-a_4 b_3-a_5 b_2-a_6 b_1$}\\\\
%\end{eqnarray}
%\begin{eqnarray}
\bf{2_s}\sim\left(
\begin{array}{c}
 a_1 b_4+a_2 b_6+a_3 b_5+a_4 b_1+a_5 b_3+a_6 b_2 \\
 a_1 b_5+a_2 b_4+a_3 b_6+a_4 b_2+a_5 b_1+a_6 b_3
\end{array}
\right)\\\\
%\end{eqnarray}
%\begin{eqnarray}
\bf{2_a}\sim\left(
\begin{array}{c}
 a_1 b_4+a_2 b_6+a_3 b_5-a_4 b_1-a_5 b_3-a_6 b_2 \\
 -a_1 b_5-a_2 b_4-a_3 b_6+a_4 b_2+a_5 b_1+a_6 b_3
\end{array}
\right)\\\\
%\end{eqnarray}
%\begin{eqnarray}
\bf{3_a}\sim\left(
\begin{array}{c}
 a_1 b_5+\omega ^2a_2 b_4+\omega a_3 b_6-\omega ^2a_4 b_2  -a_5 b_1 -\omega a_6 b_3    \\
 \omega ^2a_1 b_6 +\omega a_2 b_5+a_3 b_4-a_4 b_3 -\omega a_5 b_2  -\omega ^2a_6 b_1  \\
 \omega a_1 b_4+a_2 b_6+\omega ^2a_3 b_5-\omega a_4 b_1-\omega ^2a_5 b_3 -a_6 b_2
\end{array}
\right)\\\\
%\end{eqnarray}
%\begin{eqnarray}
\bf{\widetilde{3}_s}\sim \left(
\begin{array}{c}
 a_1 b_2+a_2 b_1+a_3 b_3-a_4 b_5-a_5 b_4-a_6 b_6 \\
 a_1 b_3+a_2 b_2+a_3 b_1-a_4 b_6-a_5 b_5-a_6 b_4 \\
 a_1 b_1+a_2 b_3+a_3 b_2-a_4 b_4-a_5 b_6-a_6 b_5
\end{array}
\right)\\\\
%\end{eqnarray}
%\begin{eqnarray}
\bf{\overline{3}_a}\sim
\left(
\begin{array}{c}
 a_1 b_4+\omega a_2 b_6+\omega ^2a_3 b_5-a_4 b_1-\omega ^2a_5 b_3 -\omega a_6 b_2     \\
 \omega ^2a_1 b_6+a_2 b_5+\omega a_3 b_4 -\omega a_4 b_3 -a_5 b_2-\omega ^2a_6 b_1     \\
 \omega a_1 b_5+\omega ^2a_2 b_4+a_3 b_6-\omega ^2a_4 b_2  -\omega a_5 b_1    -a_6 b_3
\end{array}
\right)\\\\
%\end{eqnarray}
%\begin{eqnarray}
\bf{3^{\prime}_s}\sim
\left(
\begin{array}{c}
 a_1 b_5+\omega ^2a_2 b_4+\omega a_3 b_6+\omega ^2a_4 b_2 +a_5 b_1 +\omega a_6 b_3    \\
 \omega ^2a_1 b_6 +\omega a_2 b_5+a_3 b_4+a_4 b_3 +\omega a_5 b_2  +\omega ^2a_6 b_1  \\
 \omega a_1 b_4+a_2 b_6+\omega ^2a_3 b_5+\omega a_4 b_1+\omega ^2a_5 b_3 +a_6 b_2
\end{array}
\right)\\\\
%\end{eqnarray}
%\begin{eqnarray}
\bf{\widetilde{3}^{\prime}_s}\sim
\left(
\begin{array}{c}
 a_1 b_2+a_2 b_1+a_3 b_3+a_4 b_5+a_5 b_4+a_6 b_6 \\
 a_1 b_3+a_2 b_2+a_3 b_1+a_4 b_6+a_5 b_5+a_6 b_4 \\
 a_1 b_1+a_2 b_3+a_3 b_2+a_4 b_4+a_5 b_6+a_6 b_5
\end{array}
\right)\\\\
%\end{eqnarray}
%\begin{eqnarray}
\bf{\overline{3}^{\prime}_s}\sim\left(
\begin{array}{c}
 a_1 b_4+\omega a_2 b_6+\omega ^2a_3 b_5+a_4 b_1+\omega ^2a_5 b_3 +\omega a_6 b_2     \\
 \omega ^2a_1 b_6+a_2 b_5+\omega a_3 b_4 +\omega a_4 b_3 +a_5 b_2+\omega ^2a_6 b_1     \\
 \omega a_1 b_5+\omega ^2a_2 b_4+a_3 b_6+\omega ^2a_4 b_2 +\omega a_5 b_1  +a_6 b_3
\end{array}
\right)\\
%\end{eqnarray}
%\begin{eqnarray}
\end{array}
\end{eqnarray}
\begin{eqnarray}\nonumber
\begin{array}{ccc}
\bf{6_s}\sim\left(
\begin{array}{c}
 a_4 b_5+a_5 b_4-2 a_6 b_6 \\
 a_4 b_6-2 a_5 b_5+a_6 b_4 \\
 -2 a_4 b_4+a_5 b_6+a_6 b_5 \\
 -a_1 b_2-a_2 b_1+2 a_3 b_3 \\
 -a_1 b_3+2 a_2 b_2-a_3 b_1 \\
 2 a_1 b_1-a_2 b_3-a_3 b_2
\end{array}
\right)&~~~&
%\end{eqnarray}
%\begin{eqnarray}
\bf{6_a}\sim\left(
\begin{array}{c}
 a_4 b_5-a_5 b_4 \\
 a_6 b_4-a_4 b_6 \\
 a_5 b_6-a_6 b_5 \\
 a_1 b_2-a_2 b_1 \\
 a_3 b_1-a_1 b_3 \\
 a_2 b_3-a_3 b_2
\end{array}
\right)\\\\
\end{array}
\end{eqnarray}
\end{minipage}}
\end{centering}

\end{document}